\documentclass[prl,longbibliography, aps,10pt,superscriptaddress,twocolumn,floatfix]{revtex4-1}
\usepackage{graphicx,color}
\usepackage{amsmath,amssymb,bm}
\usepackage{braket}

\usepackage[plainpages=false,pdfpagelabels,colorlinks=true,linkcolor=red,urlcolor=blue,citecolor=blue,pdftitle={Titl}e,pdfauthor={},pdfdisplaydoctitle=true,pdfduplex=DuplexFlipLongEdge]{hyperref}

\begin{document}
\title{Complexity transition in the Dicke model of light-matter interaction}
\author{Yicheng Zhang}
\author{Erhai Zhao}
\affiliation{Department of Physics and Astronomy, George Mason University, Fairfax, Virginia 22030, USA}
\date{\today}

\begin{abstract}    
Tuning the coupling strength $g$ of an interacting quantum system may drive a sudden change in its ground-state or thermal properties. To identify and grasp non-analytical, or even discontinuous, transitions in far-from-equilibrium dynamics proves more challenging. Recently Krylov complexity $C_K$ has offered fresh insights about operator growth, thermalization, and chaos in quantum dynamics. Yet it remains unclear if, and how, changing $g$ can trigger a sharp transition in the complexity measures. Here we present evidence for such a transition by mapping out the complexity phase diagram of the paradigmatic Dicke model describing two-level atoms coupled to a cavity photon mode. Two qualitatively different regimes of dynamics are identified and characterized. At the transition, the slope of $C_K$ changes suddenly to coincide with a jump in the Krylov entropy. We elucidate the nature of the regime change from the wave packet dynamics in Krylov space, where a particle is confined by a roughly linear potential but hops as if it lives in a Rindler reference frame. The competition between confinement, which leads to bouncing, and deconfinement by Rindler hopping, which leads to the destruction of wave packet analogous to gravitational spaghettification, is sensitive to the disorder in Lanczos coefficients. The framework outlined here can be applied to other quantum many-body systems.
\end{abstract}

\maketitle

Experiments on cold atoms, trapped ions, and superconducting qubits have revealed a host of novel phenomena in the far-from-equilibrium dynamics of interacting quantum many-body systems. In particular, quench dynamics, i.e. the unitary time evolution of an isolated system following a sudden change in one of the control parameters, has attracted significant interest \cite{Zhang2017Observation,Jurcevic2017Direct,Smale2019Observation,Yang2019Observation,Tian2020Observation,Muniz2020Exploring,Karch2025Probing,Bullock2026Quantum}. The notion of dynamical phase transition (DPT) was introduced to delineate qualitatively different regimes of quench dynamics by using the long-time average of some chosen observable as the ``order parameter" \cite{Marino2022Dynamical,Mori2018Thermalization,Heyl2018Dynamical,Zunkovic2018Dynamical}. The success of this proposal hinges on judicial choice of the observable based on the details of the Hamiltonian and initial state, and therefore is not always guaranteed. For instance, it has been noted that the time-averaged observables may become noisy and inconclusive when the system is close to or within a chaotic regime \cite{Lerose2018Chaotic,LewisSwan2021Characterizing,Bullock2026Quantum}. A complementary, more general approach is desired.

Recently, many-body quench dynamics has been examined fruitfully through the lens of Krylov complexity~\cite{Parker2019Universal,Balasubramanian2022Quantum,rabinovici2025krylovcomplexity,Nandy2025KrylovReview}. Certain time evolutions are more complex than others and require more computational resources to simulate. Krylov complexity $C_K(t)$ has a deep connection with circuit complexity~\cite{Craps2024Relation, Lv2024Building} and it measures the spread of the initial state in a Hilbert space with a basis constructed from Lanczos recursion. This general procedure reduces the complicated many-body dynamics to a single-particle hopping problem. Previous work has established $C_K(t)$ as a powerful diagnostic of integrable versus chaotic dynamics through the differences in early-time growth, existence of peak, and late-time plateaus~\cite{rabinovici2025krylovcomplexity,Nandy2025KrylovReview}. Moreover, $\overline{C}_K$, the time average of $C_K(t)$, was proposed as an ``order parameter" to characterize the integrability-to-chaos transition, e.g., in spin chains and random matrix theory models~\cite{Rabinovici2022Krylov,Espanol2023Assessing,Balasubramanian2025Quantum,Scialchi2024Integrability,Baggioli2025Krylov,Camargo2024Spread,Scialchi2025Exploring}, or as a marker for DPT, e.g., in the integrable Lipkin–Meshkov–Glick (LMG) model~\cite{Bento2024Krylov}. While suggestive, these results {\it did not} find nonanalytic behaviors of $\overline{C}_K$ away from integrability. Thus, it remains an {\it open conjecture} if Krylov complexity can effectively identify and characterize the transitions between distinctive regimes of quantum dynamics in realistic physical systems. 

In this letter, we present numerical evidence and analysis that strongly support this conjecture. Our main results differ from previous work in several aspects. (1) We obtain the global complexity phase diagram for the Dicke model, a paradigmatic model of light-matter interaction of fundamental interest to cavity QED and circuit QED platforms of quantum simulation and quantum control. The model does not contain random disorder but features both integrable and chaotic limits as $g$, the strength of light-matter coupling, is tuned. Its Hilbert space is infinite, which distinguishes it from finite spin chains. These make it an excellent candidate to explore complexity. (2) We observe, for the first time, non-monotonic and sharp (hence nonanalytic) changes in $\overline{C}_K$ and Krylov entropy as function of $g$, which clearly identify the complexity transition. (3) Distinctive power-law scaling of $\overline{C}_K$ with $N$, the number of spins, emerges near the transition. (4) We clarify the nature of the complexity transition by characterizing the linear asymptotics, gravitational analogue, and the roles of diagonal and off-diagonal disorder regarding the wave packet dynamics in Krylov space. 

{\it Model}---Our work is directly motivated by the realization of Dicke model \cite{Dicke1954Coherence} in trapped ion arrays \cite{Safavi2018Verification,Cohn2018Bang} and cold-atoms cavity QED \cite{Baumann2010Dicke,Klinder2015Dynamical,Zhang2018Dicke,Kroeze2018Spinor}, and the recent successful observation of DPT~\cite{Bullock2026Quantum}. The model describes the coupling between $N$ spins (two-level atoms) to a bosonic (cavity photon) mode with the following Hamiltonian ($\hbar=1$),
\begin{equation}\label{eq:dickemodel}
    \hat H=-\frac{2g}{\sqrt{N}}(\hat a+\hat a^{\dagger})\hat S_x+\delta\hat a^{\dagger}\hat a+\Omega \hat S_{z}\,.
\end{equation}
Here $\hat a^{\dagger}$ ($\hat a$) creates (annihilates) a photon with energy $\delta$, $\hat S_\alpha=1/2\sum_{j=1}^{N}\hat \sigma_j^\alpha$ are the collective spin operators defined in terms of the Pauli operators $\hat\sigma_j^\alpha$ for individual spin $j$ and component $\alpha\in\{x, y, z\}$. $\Omega$ is the energy difference between the two spin states. The ground state features a superradiant transition when the spin-boson coupling $g=\sqrt{\Omega\delta}/2$~\cite{Emary2003Chaos,Emary2023Quantum}. We introduce a dimensionless coupling strength $\tilde g=2g/\sqrt{\Omega\delta}$. For finite $N$ and at large enough $\tilde g$, the Dicke model exhibits quantum chaos which can be diagnosed by spectral statistics and out-of-time-order correlators \cite{Emary2003Chaos,Emary2023Quantum,ChavezCarlos2016Classical,LewisSwan2019Scrambling,PilatowskyCameo2020Positive,Villasenor2023Chaos,villasenor2026classical}. In the limit $\tilde g\to0$, $\hat{H}$ simplifies to the integrable Tavis-Cummings (TC) model \cite{Tavis1968Exact}, $\hat H_{\rm TC}=-({g}/{\sqrt{N}})(\hat a\hat S^++\hat a^{\dagger}\hat S^-)$, where the total number of excitations, $\hat S_z+\hat a^\dagger\hat a$, is conserved. For $\Omega\ll\delta$, after integrating out the bosons, $\hat{H}$ reduces to the LMG Hamiltonian for the collective spin only \cite{LMG1965,SafaviNaini2017dicke}. Our main interest is the time evolution of an initial product state $\ket{\psi_0}=\ket{\theta_0,\phi_0\rangle\otimes|\alpha_0}$ generated by $\hat{H}$. Here $\ket{\theta_0,\phi_0}$ is a spin coherent state pointing along the $(\theta_0,\phi_0)$ directions on the Bloch sphere, and $\ket{\alpha_0}$ is a bosonic coherent state such that $\langle \hat{a}^{\dagger} \hat{a} \rangle = \vert \alpha_0 \vert^2$. This type of initial state can be prepared in current experiments and has been used for exploring DPT and quantum-enhanced metrology \cite{Gilmore2021Harnessing,LewisSwan2021Characterizing,Zhang2025Harnessing,Bullock2026Quantum}. In particular, we will focus on $\theta_0=0$, a classical saddle point \cite{Zhang2025Harnessing}, and the photon vacuum state $\alpha_0=0$ in the main text.

{\it Krylov complexity}---We briefly summarize the state formalism of Krylov complexity to establish notation~\cite{Balasubramanian2022Quantum}. An efficient way to describe the spread of an initial state $|\psi_0\rangle$ after a quench is to apply the Gram-Schmidt orthogonalization to $\{\hat H^k|\psi_0\rangle\}$ using the recursive Lanczos algorithm to build an orthogonal Krylov basis $\{|{k}\rangle,\, k=0,1,2,\cdots\}$ \cite{Lanczos1950, viswanath_mueller_1994}. In the new basis, the Hamiltonian takes a simple tight-binding form,
\begin{equation}\label{eq:K1D}
    \hat H|{k}\rangle=a_k|k\rangle+b_k|{k-1}\rangle+b_{k+1}|{k+1}\rangle\,.
\end{equation}
Here the onsite potential $a_k$ and the hopping amplitude $b_k$ are defined as $a_k=\langle k|\hat H|k\rangle$, $b_0=0$, $b_{k>0}=\langle A_k|A_k\rangle^{1/2}$ with $|A_k\rangle= (H-a_{k-1})|k-1\rangle - b_{k-1}|k-2\rangle$, and $|k\rangle=|A_k\rangle/b_k$ with $|k=0\rangle=|\psi_0\rangle$. The time evolution of $|\psi(t)\rangle=e^{-i\hat H t}|\psi_0\rangle$ is now encoded in the wave packet dynamics of a single particle, initially located at site $k=0$, along the Krylov chain with site indices $\{k\}$. The probability of finding the particle at site $k$ and at time $t$ is then $p_k(t)=|\langle k|\psi(t)\rangle|^2$. The Krylov complexity $C_K$ is defined as the center of mass of the wave packet~\cite{Balasubramanian2022Quantum}, $C_K(t)=\sum_{k}kp_k(t)$. And the Krylov entropy $S_K(t)=-\sum_kp_k(t)\ln p_k(t)$ is the Shannon entropy for the probability distribution $p_k$. Roughly speaking, if the wave spreads far and wide on the Krylov lattice to yield large $C_K$ and $S_K$, the dynamics is more complex. The long-time average of $C_K(t)$ is denoted by $\overline{C}_K$.

\begin{figure}[!bt]
\includegraphics[width=1\columnwidth]{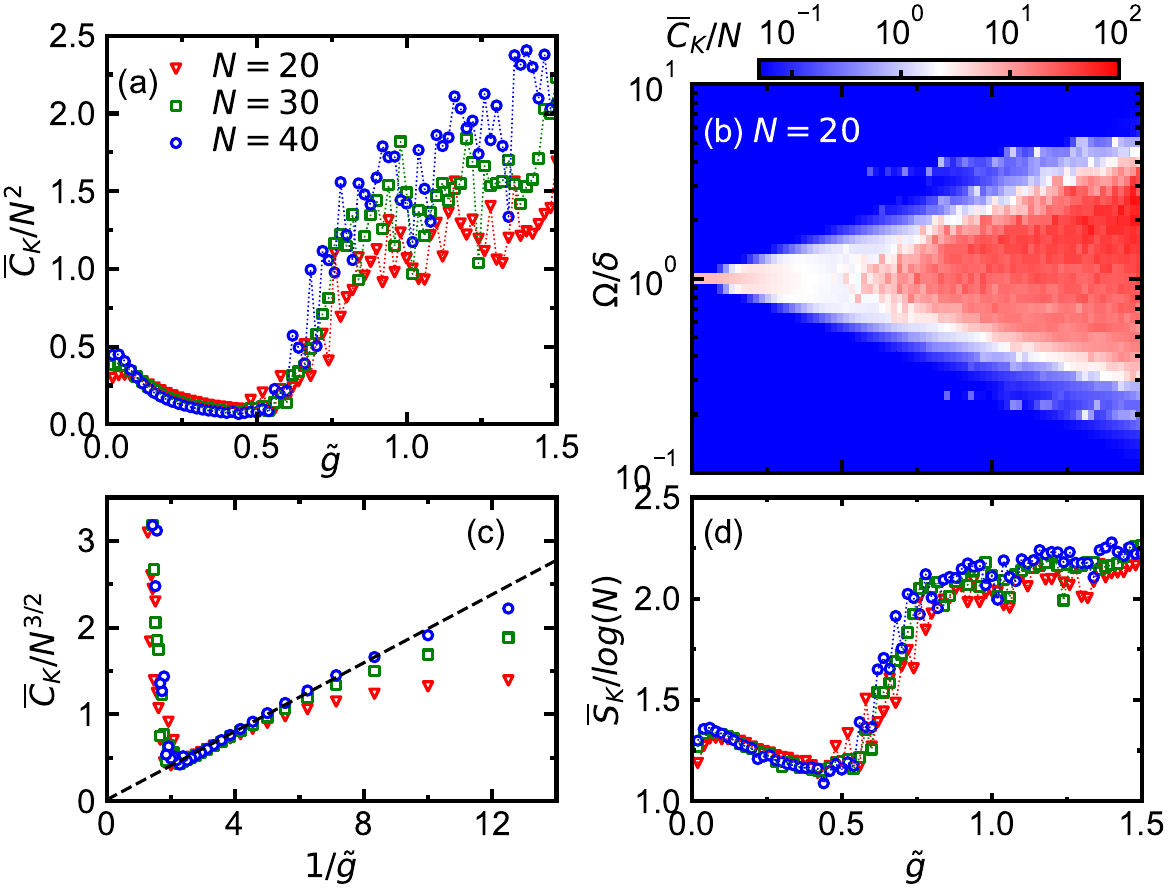}
    \caption{The Krylov dynamics of Dicke model from the initial state $\theta_0=0$ and $\alpha_0=0$. 
    (a) The long-time average of Krylov complexity $\overline{C}_K/N^2$ as a function of $\tilde g$ at resonance $\delta=\Omega$. The number of spins $N=20$ (triangles), 30 (squares), 40 (circles). (b) The color-coded
    $\overline{C}_K$ value on the plane of $\tilde g$ and $\Omega/\delta$ for $N=20$. (c) The same $\overline{C}_K$ as in (a) but plotted against $1/\tilde{g}$ to test its scaling with $N^{3/2}$.    
    (e) The Krylov entropy $\overline{S}_K$ for $\delta=\Omega$. The time average is taken over $gt\in[50,200]$.}
    \label{fig:ckaverage}
\end{figure}

\textit{Transition in complexity}---To test the conjecture, we first examine the resonance case $\delta=\Omega$ by numerically computing $C_K$ for varying $\tilde g$. In the initial state, all spins are up ($S_z=N/2$) and there is no photon. Finite $\tilde g$ creates photon excitations and introduces quantum fluctuations to $S_z$. It is reasonable to expect that the dynamics will simply become more complex as the coupling $\tilde g$ gets stronger. Thus, the $\overline{C}_K$ data plotted in Fig.~\ref{fig:ckaverage}(a) come as a surprise: the Krylov complexity first decreases as a smooth function of $\tilde g$, this trend is then disrupted around $\tilde g_c\sim 0.5$, after which $\overline{C}_K$ rises rapidly to much larger values with strong fluctuations (i.e. no longer smooth with $\tilde g$). This abrupt change is observed consistently for different system sizes, e.g. $N=20$, 30, and 40 in Fig.~\ref{fig:ckaverage}(a), with the transition getting sharper for larger $N$. Fig.~\ref{fig:ckaverage}(c) shows that the data points from different $N$'s collapse onto a single curve (dashed line) to suggest $\overline C_K\sim N^{3/2}$ as the transition is approached from $\tilde{g}<\tilde{g}_c$. For $\tilde{g}>\tilde{g}_c$ away from the transition, the data are consistent with $\overline C_K\sim N^{5/2}$ instead~\cite{SM}. Together, the sharp change in the slope of $\overline C_K(\tilde g)$ and the emergence of a distinctive $N^{3/2}$ scaling identify a robust complexity transition between two regimes of quench dynamics.

Further evidence for the transition comes from the time-averaged Krylov entropy $\overline{S}_K$ shown in Fig.~\ref{fig:ckaverage}(d). As a function of $\tilde g$ it follows a pattern similar to $\overline{C}_K$. The $\tilde{g}<\tilde{g}_c$ region on the left has lower entropy, i.e. the wave packet is more localized on the Krylov lattice. The $\tilde{g}>\tilde{g}_c$ region has much larger entropy characteristic of an extended waveform. The rapid rise of entropy across the transition is reminiscent of a first-order transition where the entropy jump is responsible for the latent heat. The complexity transition is not a special feature limited to $\Omega=\delta$ or the initial state with $\theta_0=0$. Fig.~\ref{fig:ckaverage}(b) shows the complexity phase diagram of $\overline{C}_K$ (in log scale) on the plane of $\tilde g$ and $\Omega/\delta$. Figs.~\ref{fig:ckaverage}(a) and (d) then correspond to a horizontal cut at $\Omega/\delta=1$ across Fig.~\ref{fig:ckaverage}(b). The blue region has negligible $\overline{C}_K$, where the spin is locked near its initial value with no bosonic excitations. Additional data for other initial states and $\Omega\neq\delta$ are provided in Fig.~\ref{fig:otherconditions} of End Matter.

\textit{Identifying the two regimes}---The complexity phase diagram demonstrates the utility of $C_K$ as an independent probe of quantum dynamics. It serves as a starting point to identify the two regimes and investigate if the transition coincides with any DPT, or 
the onset of thermalization or chaos. Note that there is no a priori reason for these changes to be concurrent. In previous work \cite{LewisSwan2021Characterizing, Bullock2026Quantum}, the long-time average of $\hat{S}_z$ and boson number $\hat{n}_b=\hat{a}^\dagger \hat{a}$ were used as order parameters to diagnose DPT in the Dicke model (but for different initial states). As seen from Fig.~\ref{fig:transition} (a), both averages stay roughly flat for $\tilde{g}<\tilde{g}_c$, and beyond $\tilde{g}_c$, $\overline{\langle\hat S_z\rangle}$ decreases to approach zero while $\overline{\langle\hat n_a\rangle}$ increases. There is no symmetry breaking at $g_c\sim 0.5$ which is away from the critical points of zero- or finite-temperature superradiant transitions. Both $\langle\hat S_z(t)\rangle$ and $\langle\hat n_b(t)\rangle$ exhibit regular oscillations with time for $\tilde{g}<\tilde{g_c}$~\cite{SM}. The oscillation is replaced by seemingly chaotic time evolution for $\tilde{g}>\tilde{g}_c$. These observations prompt us to refer to the two regimes as ``regular" and ``chaotic", respectively.

In End Matter, we present evidence that the complexity transition observed here is broadly aligned with the onset of chaos. The oscillatory dynamics in the regular regime retains the characteristics of the integrable Tavis-Cummings limit: $|\psi(t)\rangle$ maintains a large overlap with the TC sector \cite{SM}, while the energy level statistics (within an interval around the initial energy) is consistent with the Poisson distribution of integrable systems. In contrast, level statistics deep inside the chaotic regime is consistent with the Wigner-Dyson distribution of chaotic Hamiltonians, and the majority of semiclassical trajectories in the phase space \cite{Chavez2016Classical,LewisSwan2019Scrambling,Skokos2010Book} acquire finite Lyapunov exponents at $\tilde{g}>\tilde{g}_c$. Note that these integrability/chaos measures, while highly suggestive when used in tandem, are either indirectly tied to the full quantum dynamics of a given initial state or do not yield a clear, consistent ``phase boundary." The complexity measures $C_K$ and $S_K$ characterize $|\psi(t)\rangle$ directly. Moreover, as we shall show below, reducing the many-body dynamics to a hopping problem with $(a_k,b_k)$ enables an intuitive, deeper understanding of the two regimes and the transition.

\begin{figure}[!bt]
\includegraphics[width=1\columnwidth]{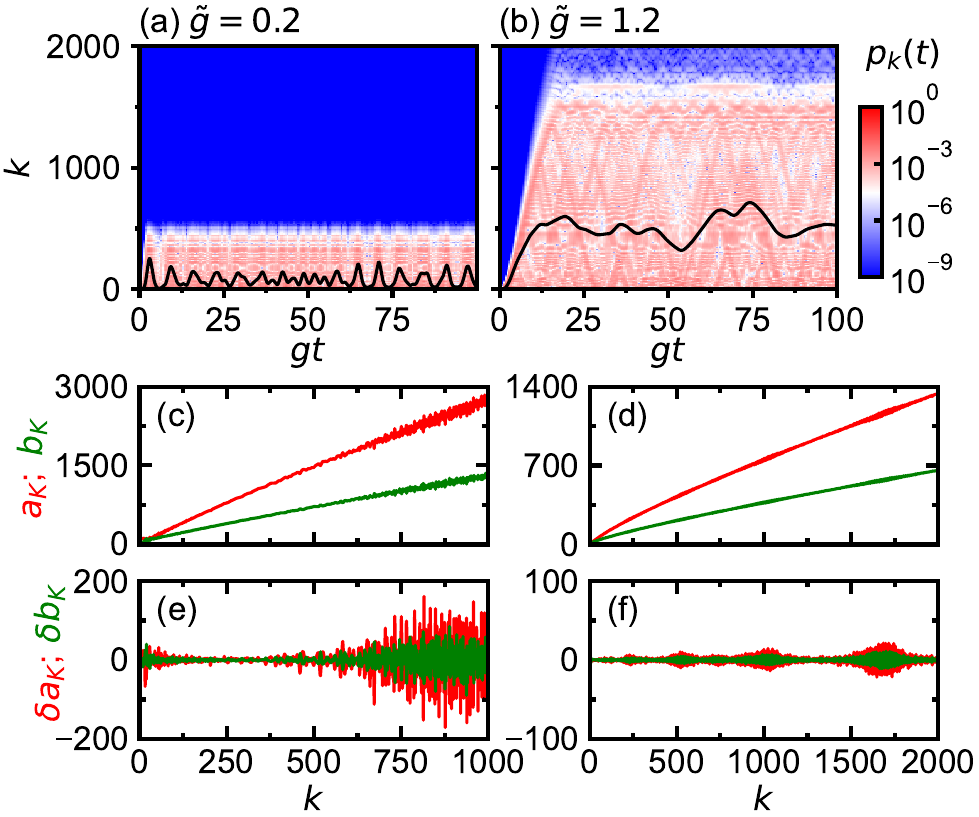}
    \caption{Left column: (a) The probability $p_k(t)$ of finding the particle on site $k$ of the Krylov lattice at time $t$ for $\tilde g=0.2$ in the regular regime. The solid line shows $C_K(t)$. (c) The corresponding $a_K$ (red) and $b_K$ (green) are roughly linear in $k$. (e) The profiles of diagonal $\delta a_K$ (red) and off-diagonal $\delta b_K$ (green) disorder, where $\overline{a}_K$ and $\overline{b}_K$ are sliding self-averages over $k\pm10$. Right column: (b), (d), (f) are similar plots for $\tilde g=1.2$ in the chaotic regime. $N=20$ and $\delta=\Omega$.}
    \label{fig:krylovlattice}
\end{figure}

\textit{Breathing versus spaghettification in Krylov space}---Figs. \ref{fig:krylovlattice}(c)-(d) show that overall $a_k$ and $b_k$ both increase linearly with $k$, except in a small region $k\lesssim 25$. To capture the asymptotic wave dynamics, we first consider a toy model, where $a_k$ and $b_k$ are strictly linear, 
\begin{equation}\label{linearab}
   a_k=k,\quad b_k=\eta k,\quad k=0,1,2,...  
\end{equation}
We have set $a_1$ as the energy unit, which leaves $\eta=b_k/a_k$ as the only tuning parameter. The onsite term $a_k$ represents a linear confining potential $V(x>0)=x$, while the position-dependent hopping $b_k$ mimics a nonuniform spacetime metric~\cite{rodriguez2017synthetic,morice2021synthetic,morice2022quantum,lv2022curving,analog-unruh}, i.e. synthetic gravity~\cite{barcelo2011analogue}. More specifically, $b_k=\eta k$ describes the (1+1)D Rindler spacetime experienced by an observer moving along the $x$ axis with constant acceleration $\eta$~\cite{rindler,misner1973gravitation}. The local clock at larger $x$ ticks faster, $d\tau = (\eta x) dt$, hence a tunneling rate linear to $x$. Alternatively, via the equivalence principle, the accelerating observer experiences a gravitational pull towards the origin $k=0$, so that an outgoing wave packet becomes elongated due to the continuous redshift.

The Krylov dynamics then describes a particle initially at $k=0$ whose fate depends on $\eta$, i.e., the competition between confinement by $a_k$ and delocalization by Rinder hopping $b_k$. For $\eta<1/2$, one can show that the wave packet disperses but stays close to $k=0$ to breath: $\langle k|\psi(t)\rangle\propto e^{-k/w(t)}$ where the width $w(t)$ is a periodic function of $t$ with an amplitude increasing with $\eta$. In other words, its center of mass $C_K(t)$ oscillates, analogous to a bouncing ball. At the critical point $\eta=0.5$, a delocalization transition occurs. Here the eigenstates $\phi_E(k)$ for low energy $E$ are no longer localized near $k=0$. Instead, they exhibit power law decay. For a finite open chain, we find $\phi_E(k) = (-1)^k \,_{1}F_1 (-k;1;2E+1)$ where $_{1}F_1$ is the confluent hypergeometric function. For $k\gg 1$, $\phi_E(k) \propto (-1)^k k^{-1/4}\cos (2\sqrt{(2E+1)k}-\pi/4)$, where the oscillation shows chirping due to the continuous red shift. The wave packet continues to flatten and never comes back to refocus at $k=0$. The extreme stretching is reminiscent of spaghettification near a black hole~\cite{hawking2009brief,pinochet2022little}. For the open chain, after reaching the boundary $k=L$, the wave reduces to ripples to fill the entire lattice. Accordingly, $C_K(t)$ rises to $L/2$ then decays slowly, see Figs. \ref{fig:lanczossmooth}(a)-(b). The oscillations of $C_K(t)$ and $S_K(t)$ are gone. Instead, the hallmark of spaghettification is a large $\overline{S}_K$.

The asymptotic linear increase of $b_k$ was key to the universal operator growth hypothesis \cite{Parker2019Universal}.Linear $a_k$ and $b_k$ were noted for harmonic oscillator models \cite{Balasubramanian2022Quantum}\footnote{In Ref.~\cite{Balasubramanian2022Quantum}, the transition is at $\omega=0$ from the standard to the inverted oscillator, which are connected by analytical continuation $\omega \rightarrow -i\omega$. The model and motivation are different from our case.}. But the specific gravitational analog and the fact that $\eta$ drives a delocalization transition have not been elucidated before. Despite being a crude approximation of the actual $(a_k,b_k)$ data, the linear model Eq. \eqref{linearab} is able to capture the limits of oscillating wave packets and their destruction by spaghettification. For the Dicke model at resonance, the $\eta$ value extracted from the linear fit increases monotonically with $\tilde{g}$ to approach $0.5$ from below, so it is not far from the delocalization transition \cite{SM}. 

\textit{Confinement by disorder}---Deviations of $(a_k,b_k)$ from the linear fit occur in two places: (A) for $k\lesssim 25$, $a_k$ stays relatively flat while $b_k$ increases with a larger slope; (B) within certain intervals of $k$, the seemingly random site-to-site fluctuations become more pronounced. To separate these two effects, we consider a ``smoothed model" where each $a_k$ ($b_k$) is replaced by $\overline{a}_k$ ($\overline{b}_k$), the sliding average of ten neighboring values, which effectively eliminates (B). Define the diagonal (off-diagonal) disorder as $\delta a_k=a_k-\overline{a}_k$ ($\delta b_k=b_k-\overline{b}_k$). Figs.~\ref{fig:krylovlattice}(e)-(f) give two examples of the $(\delta a_k, \delta b_k)$ profile showing clearly visible local peaks. The main effect of (A) is to launch the wave packet, i.e., the smoothed model produces a wave packet leaving the small $k$ region with finite momentum, in contrast to the breathing behavior of the linear model above. Despite this difference, at small $\tilde{g}$, the dispersing wave packet still gets reflected by the strong confinement, resulting in an oscillatory $C_K(t)$. For large $\tilde{g}$, as the wave packet travels further to the right and slows down by the confinement, it is continuously stretched by Rindler hopping $b_k$ and shreds weight from its tail. Eventually it disintegrates into ripples and flares to fill a long stretch of the $k$ axis, $k\in[0,k_{max}]$, and cannot refocus. The oscillation is preempted by spaghettification.

Now we are in position to pin down the effects of disorder $(\delta a_k, \delta b_k)$ to obtain a complete, accurate picture of the Krylov dynamics. (1) Scattering by the disorder slows down the speed of the wave packet. (2) $\delta a_k$ induces phase shifts, while $\delta b_k$ leads to fluctuations in the group velocity, so the wave shape is distorted in regions of strong disorder. (3) The disorder landscape $(\delta a_k, \delta b_k)$ varies with $\tilde{g}$ to restricts $k_{max}$, the range of the wave packet [the boundary of the pink region in Fig. \ref{fig:krylovlattice}(a)-(b)]. For example, at $\tilde{g}=0.2$, beyond the launching region, $\delta a_k$ (and similarly for $b_k$) remains small until it rises rapidly around $k=500$ before reaching its first peak, see Fig. \ref{fig:krylovlattice}(e). The peak acts as a barrier to confine the wave within a range $k_{max}\sim 450$. As $\tilde{g}$ increases, this disorder peak moves towards $k=0$ so the oscillating wave packet becomes {increasingly confined}, which explains the downward shift in $\overline{C}_K$ observed in Fig. \ref{fig:ckaverage}(a). This trend stops at $\tilde{g}_c$ where the first peak starts to overlap with the launching region. A rather different disorder landscape emerges for $\tilde{g}>\tilde{g}_c$: it comprises a series of small, broad bumps [Fig. \ref{fig:krylovlattice}(f)] which act as weak perturbations to slow down and scramble the wave. In this regime, $k_{max}$ measures the extent of spaghettification rather than the amplitude of oscillations.

To summarize, the ``regular" regime is distinguished by confined, oscillating wave packets, thanks to relatively small Rindler acceleration $\eta$ and strong nonuniform disorder $(\delta a_k, \delta b_k)$, while the ``chaotic" regime is characterized by spaghettification, i.e. the destruction of wave packets by near-critical $\eta$. The critical point $g_c$ marks the maximum confinement of the wave packet with $k_{max}\sim 200$. Disorder plays a decisive role, without which $\overline{C}_K$ and $\overline{S}_K$ would increase monotonically with $\tilde{g}$, see Figs. \ref{fig:lanczossmooth}(c)-(d).

{\it Outlook}---It is natural to expect that dynamics gets more complex (to involve larger portions of the Hilbert space and demand more resources to simulate) at intermediate coupling, e.g., close to quantum criticality between perturbative limits. Our results for the Dicke model show that this can be false: $\overline C_K$ actually achieves a minimum at $\tilde{g}\sim \tilde{g}_c$. More surprisingly, $\overline C_K$ and $\overline S_K$ show nonanalytical, sharp changes in slope. The existence of qualitatively different regimes of quench dynamics is also corroborated by the distinct scaling behaviors of $\overline C_K$ with $N$. Thus, Krylov spread complexity offers an independent, effective probe to identify distinct dynamic regimes. It complements traditional measures such as time-averaged observables and level statistics which may not yield clean-cut signals due to enhanced fluctuations. Fig.~\ref{fig:ckaverage} highlights the utility of this approach.

Recent trapped ion experiments have realized Dicke model with $N$ up to 100 and observed DPT in the integrable limit, and they are poised to explore the many-body dynamics for general ${g}$ and $\Omega/\delta$, including the transition into chaos \cite{Bullock2026Quantum}. Fig. \ref{fig:ckaverage}(b) can serve as a map to aid future experiments to better understand the interplay between complexity, thermalization, and chaos. This program of locating complexity transitions can be applied to any Hamiltonian or initial state. It also offers a clear pathway to gain physical understanding about the dynamic regimes and transitions in terms of wave packet dynamics in the presence of synthetic gravity. It would be exciting to explore what kinds of $(a_k, b_k)$ other Hamiltonians lead to, and how they differ from the oscillation versus spaghettification observed here.

This work is supported by AFOSR Grant FA9550-23-1-0598 and NSF Grant PHY-206419. Y.Z. thank Robert Lewis-Swan for insightful discussion.

\bibliographystyle{apsrev4-1}
\bibliography{Reference}

\begin{thebibliography}{65}%
\makeatletter
\providecommand \@ifxundefined [1]{%
 \@ifx{#1\undefined}
}%
\providecommand \@ifnum [1]{%
 \ifnum #1\expandafter \@firstoftwo
 \else \expandafter \@secondoftwo
 \fi
}%
\providecommand \@ifx [1]{%
 \ifx #1\expandafter \@firstoftwo
 \else \expandafter \@secondoftwo
 \fi
}%
\providecommand \natexlab [1]{#1}%
\providecommand \enquote  [1]{``#1''}%
\providecommand \bibnamefont  [1]{#1}%
\providecommand \bibfnamefont [1]{#1}%
\providecommand \citenamefont [1]{#1}%
\providecommand \href@noop [0]{\@secondoftwo}%
\providecommand \href [0]{\begingroup \@sanitize@url \@href}%
\providecommand \@href[1]{\@@startlink{#1}\@@href}%
\providecommand \@@href[1]{\endgroup#1\@@endlink}%
\providecommand \@sanitize@url [0]{\catcode `\\12\catcode `\$12\catcode
  `\&12\catcode `\#12\catcode `\^12\catcode `\_12\catcode `\%12\relax}%
\providecommand \@@startlink[1]{}%
\providecommand \@@endlink[0]{}%
\providecommand \url  [0]{\begingroup\@sanitize@url \@url }%
\providecommand \@url [1]{\endgroup\@href {#1}{\urlprefix }}%
\providecommand \urlprefix  [0]{URL }%
\providecommand \Eprint [0]{\href }%
\providecommand \doibase [0]{http://dx.doi.org/}%
\providecommand \selectlanguage [0]{\@gobble}%
\providecommand \bibinfo  [0]{\@secondoftwo}%
\providecommand \bibfield  [0]{\@secondoftwo}%
\providecommand \translation [1]{[#1]}%
\providecommand \BibitemOpen [0]{}%
\providecommand \bibitemStop [0]{}%
\providecommand \bibitemNoStop [0]{.\EOS\space}%
\providecommand \EOS [0]{\spacefactor3000\relax}%
\providecommand \BibitemShut  [1]{\csname bibitem#1\endcsname}%
\let\auto@bib@innerbib\@empty
\bibitem [{\citenamefont {Zhang}\ \emph {et~al.}(2017)\citenamefont {Zhang},
  \citenamefont {Pagano}, \citenamefont {Hess}, \citenamefont {Kyprianidis},
  \citenamefont {Becker}, \citenamefont {Kaplan}, \citenamefont {Gorshkov},
  \citenamefont {Gong},\ and\ \citenamefont {Monroe}}]{Zhang2017Observation}%
  \BibitemOpen
  \bibfield  {author} {\bibinfo {author} {\bibfnamefont {J.}~\bibnamefont
  {Zhang}}, \bibinfo {author} {\bibfnamefont {G.}~\bibnamefont {Pagano}},
  \bibinfo {author} {\bibfnamefont {P.~W.}\ \bibnamefont {Hess}}, \bibinfo
  {author} {\bibfnamefont {A.}~\bibnamefont {Kyprianidis}}, \bibinfo {author}
  {\bibfnamefont {P.}~\bibnamefont {Becker}}, \bibinfo {author} {\bibfnamefont
  {H.}~\bibnamefont {Kaplan}}, \bibinfo {author} {\bibfnamefont {A.~V.}\
  \bibnamefont {Gorshkov}}, \bibinfo {author} {\bibfnamefont {Z.~X.}\
  \bibnamefont {Gong}}, \ and\ \bibinfo {author} {\bibfnamefont
  {C.}~\bibnamefont {Monroe}},\ }\href {\doibase 10.1038/nature24654}
  {\bibfield  {journal} {\bibinfo  {journal} {Nature}\ }\textbf {\bibinfo
  {volume} {551}},\ \bibinfo {pages} {601} (\bibinfo {year}
  {2017})}\BibitemShut {NoStop}%
\bibitem [{\citenamefont {Jurcevic}\ \emph {et~al.}(2017)\citenamefont
  {Jurcevic}, \citenamefont {Shen}, \citenamefont {Hauke}, \citenamefont
  {Maier}, \citenamefont {Brydges}, \citenamefont {Hempel}, \citenamefont
  {Lanyon}, \citenamefont {Heyl}, \citenamefont {Blatt},\ and\ \citenamefont
  {Roos}}]{Jurcevic2017Direct}%
  \BibitemOpen
  \bibfield  {author} {\bibinfo {author} {\bibfnamefont {P.}~\bibnamefont
  {Jurcevic}}, \bibinfo {author} {\bibfnamefont {H.}~\bibnamefont {Shen}},
  \bibinfo {author} {\bibfnamefont {P.}~\bibnamefont {Hauke}}, \bibinfo
  {author} {\bibfnamefont {C.}~\bibnamefont {Maier}}, \bibinfo {author}
  {\bibfnamefont {T.}~\bibnamefont {Brydges}}, \bibinfo {author} {\bibfnamefont
  {C.}~\bibnamefont {Hempel}}, \bibinfo {author} {\bibfnamefont {B.~P.}\
  \bibnamefont {Lanyon}}, \bibinfo {author} {\bibfnamefont {M.}~\bibnamefont
  {Heyl}}, \bibinfo {author} {\bibfnamefont {R.}~\bibnamefont {Blatt}}, \ and\
  \bibinfo {author} {\bibfnamefont {C.~F.}\ \bibnamefont {Roos}},\ }\href
  {\doibase 10.1103/PhysRevLett.119.080501} {\bibfield  {journal} {\bibinfo
  {journal} {Phys. Rev. Lett.}\ }\textbf {\bibinfo {volume} {119}},\ \bibinfo
  {pages} {080501} (\bibinfo {year} {2017})}\BibitemShut {NoStop}%
\bibitem [{\citenamefont {Smale}\ \emph {et~al.}(2019)\citenamefont {Smale},
  \citenamefont {He}, \citenamefont {Olsen}, \citenamefont {Jackson},
  \citenamefont {Sharum}, \citenamefont {Trotzky}, \citenamefont {Marino},
  \citenamefont {Rey},\ and\ \citenamefont {Thywissen}}]{Smale2019Observation}%
  \BibitemOpen
  \bibfield  {author} {\bibinfo {author} {\bibfnamefont {S.}~\bibnamefont
  {Smale}}, \bibinfo {author} {\bibfnamefont {P.}~\bibnamefont {He}}, \bibinfo
  {author} {\bibfnamefont {B.~A.}\ \bibnamefont {Olsen}}, \bibinfo {author}
  {\bibfnamefont {K.~G.}\ \bibnamefont {Jackson}}, \bibinfo {author}
  {\bibfnamefont {H.}~\bibnamefont {Sharum}}, \bibinfo {author} {\bibfnamefont
  {S.}~\bibnamefont {Trotzky}}, \bibinfo {author} {\bibfnamefont
  {J.}~\bibnamefont {Marino}}, \bibinfo {author} {\bibfnamefont {A.~M.}\
  \bibnamefont {Rey}}, \ and\ \bibinfo {author} {\bibfnamefont {J.~H.}\
  \bibnamefont {Thywissen}},\ }\href {\doibase 10.1126/sciadv.aax1568}
  {\bibfield  {journal} {\bibinfo  {journal} {Science Advances}\ }\textbf
  {\bibinfo {volume} {5}},\ \bibinfo {pages} {eaax1568} (\bibinfo {year}
  {2019})}\BibitemShut {NoStop}%
\bibitem [{\citenamefont {Yang}\ \emph {et~al.}(2019)\citenamefont {Yang},
  \citenamefont {Tian}, \citenamefont {Yang}, \citenamefont {Qiu},
  \citenamefont {Liang}, \citenamefont {Chu}, \citenamefont
  {Da\ifmmode~\breve{g}\else \u{g}\fi{}}, \citenamefont {Xu}, \citenamefont
  {Liu},\ and\ \citenamefont {Duan}}]{Yang2019Observation}%
  \BibitemOpen
  \bibfield  {author} {\bibinfo {author} {\bibfnamefont {H.-X.}\ \bibnamefont
  {Yang}}, \bibinfo {author} {\bibfnamefont {T.}~\bibnamefont {Tian}}, \bibinfo
  {author} {\bibfnamefont {Y.-B.}\ \bibnamefont {Yang}}, \bibinfo {author}
  {\bibfnamefont {L.-Y.}\ \bibnamefont {Qiu}}, \bibinfo {author} {\bibfnamefont
  {H.-Y.}\ \bibnamefont {Liang}}, \bibinfo {author} {\bibfnamefont {A.-J.}\
  \bibnamefont {Chu}}, \bibinfo {author} {\bibfnamefont {C.~B.}\ \bibnamefont
  {Da\ifmmode~\breve{g}\else \u{g}\fi{}}}, \bibinfo {author} {\bibfnamefont
  {Y.}~\bibnamefont {Xu}}, \bibinfo {author} {\bibfnamefont {Y.}~\bibnamefont
  {Liu}}, \ and\ \bibinfo {author} {\bibfnamefont {L.-M.}\ \bibnamefont
  {Duan}},\ }\href {\doibase 10.1103/PhysRevA.100.013622} {\bibfield  {journal}
  {\bibinfo  {journal} {Phys. Rev. A}\ }\textbf {\bibinfo {volume} {100}},\
  \bibinfo {pages} {013622} (\bibinfo {year} {2019})}\BibitemShut {NoStop}%
\bibitem [{\citenamefont {Tian}\ \emph {et~al.}(2020)\citenamefont {Tian},
  \citenamefont {Yang}, \citenamefont {Qiu}, \citenamefont {Liang},
  \citenamefont {Yang}, \citenamefont {Xu},\ and\ \citenamefont
  {Duan}}]{Tian2020Observation}%
  \BibitemOpen
  \bibfield  {author} {\bibinfo {author} {\bibfnamefont {T.}~\bibnamefont
  {Tian}}, \bibinfo {author} {\bibfnamefont {H.-X.}\ \bibnamefont {Yang}},
  \bibinfo {author} {\bibfnamefont {L.-Y.}\ \bibnamefont {Qiu}}, \bibinfo
  {author} {\bibfnamefont {H.-Y.}\ \bibnamefont {Liang}}, \bibinfo {author}
  {\bibfnamefont {Y.-B.}\ \bibnamefont {Yang}}, \bibinfo {author}
  {\bibfnamefont {Y.}~\bibnamefont {Xu}}, \ and\ \bibinfo {author}
  {\bibfnamefont {L.-M.}\ \bibnamefont {Duan}},\ }\href {\doibase
  10.1103/PhysRevLett.124.043001} {\bibfield  {journal} {\bibinfo  {journal}
  {Phys. Rev. Lett.}\ }\textbf {\bibinfo {volume} {124}},\ \bibinfo {pages}
  {043001} (\bibinfo {year} {2020})}\BibitemShut {NoStop}%
\bibitem [{\citenamefont {Muniz}\ \emph {et~al.}(2020)\citenamefont {Muniz},
  \citenamefont {Barberena}, \citenamefont {Lewis-Swan}, \citenamefont {Young},
  \citenamefont {Cline}, \citenamefont {Rey},\ and\ \citenamefont
  {Thompson}}]{Muniz2020Exploring}%
  \BibitemOpen
  \bibfield  {author} {\bibinfo {author} {\bibfnamefont {J.~A.}\ \bibnamefont
  {Muniz}}, \bibinfo {author} {\bibfnamefont {D.}~\bibnamefont {Barberena}},
  \bibinfo {author} {\bibfnamefont {R.~J.}\ \bibnamefont {Lewis-Swan}},
  \bibinfo {author} {\bibfnamefont {D.~J.}\ \bibnamefont {Young}}, \bibinfo
  {author} {\bibfnamefont {J.~R.~K.}\ \bibnamefont {Cline}}, \bibinfo {author}
  {\bibfnamefont {A.~M.}\ \bibnamefont {Rey}}, \ and\ \bibinfo {author}
  {\bibfnamefont {J.~K.}\ \bibnamefont {Thompson}},\ }\href {\doibase
  10.1038/s41586-020-2224-x} {\bibfield  {journal} {\bibinfo  {journal}
  {Nature}\ }\textbf {\bibinfo {volume} {580}},\ \bibinfo {pages} {602}
  (\bibinfo {year} {2020})}\BibitemShut {NoStop}%
\bibitem [{\citenamefont {Karch}\ \emph {et~al.}(2025)\citenamefont {Karch},
  \citenamefont {Bandyopadhyay}, \citenamefont {Sun}, \citenamefont {Impertro},
  \citenamefont {Huh}, \citenamefont {Rodríguez}, \citenamefont {Wienand},
  \citenamefont {Ketterle}, \citenamefont {Heyl}, \citenamefont {Polkovnikov},
  \citenamefont {Bloch},\ and\ \citenamefont
  {Aidelsburger}}]{Karch2025Probing}%
  \BibitemOpen
  \bibfield  {author} {\bibinfo {author} {\bibfnamefont {S.}~\bibnamefont
  {Karch}}, \bibinfo {author} {\bibfnamefont {S.}~\bibnamefont
  {Bandyopadhyay}}, \bibinfo {author} {\bibfnamefont {Z.-H.}\ \bibnamefont
  {Sun}}, \bibinfo {author} {\bibfnamefont {A.}~\bibnamefont {Impertro}},
  \bibinfo {author} {\bibfnamefont {S.}~\bibnamefont {Huh}}, \bibinfo {author}
  {\bibfnamefont {I.~P.}\ \bibnamefont {Rodríguez}}, \bibinfo {author}
  {\bibfnamefont {J.~F.}\ \bibnamefont {Wienand}}, \bibinfo {author}
  {\bibfnamefont {W.}~\bibnamefont {Ketterle}}, \bibinfo {author}
  {\bibfnamefont {M.}~\bibnamefont {Heyl}}, \bibinfo {author} {\bibfnamefont
  {A.}~\bibnamefont {Polkovnikov}}, \bibinfo {author} {\bibfnamefont
  {I.}~\bibnamefont {Bloch}}, \ and\ \bibinfo {author} {\bibfnamefont
  {M.}~\bibnamefont {Aidelsburger}},\ }\href {https://arxiv.org/abs/2501.16995}
  {\enquote {\bibinfo {title} {Probing quantum many-body dynamics using
  subsystem loschmidt echos},}\ } (\bibinfo {year} {2025}),\ \Eprint
  {http://arxiv.org/abs/2501.16995} {arXiv:2501.16995 [cond-mat.quant-gas]}
  \BibitemShut {NoStop}%
\bibitem [{\citenamefont {Bullock}\ \emph {et~al.}(2026)\citenamefont
  {Bullock}, \citenamefont {Muleady}, \citenamefont {Lilieholm}, \citenamefont
  {Zhang}, \citenamefont {Safavi-Naini}, \citenamefont {Lewis-Swan},
  \citenamefont {Bollinger}, \citenamefont {Rey},\ and\ \citenamefont
  {Carter}}]{Bullock2026Quantum}%
  \BibitemOpen
  \bibfield  {author} {\bibinfo {author} {\bibfnamefont {B.}~\bibnamefont
  {Bullock}}, \bibinfo {author} {\bibfnamefont {S.~R.}\ \bibnamefont
  {Muleady}}, \bibinfo {author} {\bibfnamefont {J.~F.}\ \bibnamefont
  {Lilieholm}}, \bibinfo {author} {\bibfnamefont {Y.}~\bibnamefont {Zhang}},
  \bibinfo {author} {\bibfnamefont {A.}~\bibnamefont {Safavi-Naini}}, \bibinfo
  {author} {\bibfnamefont {R.~J.}\ \bibnamefont {Lewis-Swan}}, \bibinfo
  {author} {\bibfnamefont {J.~J.}\ \bibnamefont {Bollinger}}, \bibinfo {author}
  {\bibfnamefont {A.~M.}\ \bibnamefont {Rey}}, \ and\ \bibinfo {author}
  {\bibfnamefont {A.~L.}\ \bibnamefont {Carter}},\ }\href
  {https://arxiv.org/abs/2602.06114} {\enquote {\bibinfo {title} {Quantum
  simulation of the dicke model in a two-dimensional ion crystal: chaos,
  quantum thermalization, and revivals},}\ } (\bibinfo {year} {2026}),\ \Eprint
  {http://arxiv.org/abs/2602.06114} {arXiv:2602.06114 [quant-ph]} \BibitemShut
  {NoStop}%
\bibitem [{\citenamefont {Marino}\ \emph {et~al.}(2022)\citenamefont {Marino},
  \citenamefont {Eckstein}, \citenamefont {Foster},\ and\ \citenamefont
  {Rey}}]{Marino2022Dynamical}%
  \BibitemOpen
  \bibfield  {author} {\bibinfo {author} {\bibfnamefont {J.}~\bibnamefont
  {Marino}}, \bibinfo {author} {\bibfnamefont {M.}~\bibnamefont {Eckstein}},
  \bibinfo {author} {\bibfnamefont {M.~S.}\ \bibnamefont {Foster}}, \ and\
  \bibinfo {author} {\bibfnamefont {A.~M.}\ \bibnamefont {Rey}},\ }\href
  {\doibase 10.1088/1361-6633/ac906c} {\bibfield  {journal} {\bibinfo
  {journal} {Reports on Progress in Physics}\ }\textbf {\bibinfo {volume}
  {85}},\ \bibinfo {pages} {116001} (\bibinfo {year} {2022})}\BibitemShut
  {NoStop}%
\bibitem [{\citenamefont {Mori}\ \emph {et~al.}(2018)\citenamefont {Mori},
  \citenamefont {Ikeda}, \citenamefont {Kaminishi},\ and\ \citenamefont
  {Ueda}}]{Mori2018Thermalization}%
  \BibitemOpen
  \bibfield  {author} {\bibinfo {author} {\bibfnamefont {T.}~\bibnamefont
  {Mori}}, \bibinfo {author} {\bibfnamefont {T.~N.}\ \bibnamefont {Ikeda}},
  \bibinfo {author} {\bibfnamefont {E.}~\bibnamefont {Kaminishi}}, \ and\
  \bibinfo {author} {\bibfnamefont {M.}~\bibnamefont {Ueda}},\ }\href {\doibase
  10.1088/1361-6455/aabcdf} {\bibfield  {journal} {\bibinfo  {journal} {Journal
  of Physics B: Atomic, Molecular and Optical Physics}\ }\textbf {\bibinfo
  {volume} {51}},\ \bibinfo {pages} {112001} (\bibinfo {year}
  {2018})}\BibitemShut {NoStop}%
\bibitem [{\citenamefont {Heyl}(2018)}]{Heyl2018Dynamical}%
  \BibitemOpen
  \bibfield  {author} {\bibinfo {author} {\bibfnamefont {M.}~\bibnamefont
  {Heyl}},\ }\href {\doibase 10.1088/1361-6633/aaaf9a} {\bibfield  {journal}
  {\bibinfo  {journal} {Reports on Progress in Physics}\ }\textbf {\bibinfo
  {volume} {81}},\ \bibinfo {pages} {054001} (\bibinfo {year}
  {2018})}\BibitemShut {NoStop}%
\bibitem [{\citenamefont {\ifmmode \check{Z}\else
  \v{Z}\fi{}unkovi\ifmmode~\check{c}\else \v{c}\fi{}}\ \emph
  {et~al.}(2018)\citenamefont {\ifmmode \check{Z}\else
  \v{Z}\fi{}unkovi\ifmmode~\check{c}\else \v{c}\fi{}}, \citenamefont {Heyl},
  \citenamefont {Knap},\ and\ \citenamefont {Silva}}]{Zunkovic2018Dynamical}%
  \BibitemOpen
  \bibfield  {author} {\bibinfo {author} {\bibfnamefont {B.}~\bibnamefont
  {\ifmmode \check{Z}\else \v{Z}\fi{}unkovi\ifmmode~\check{c}\else
  \v{c}\fi{}}}, \bibinfo {author} {\bibfnamefont {M.}~\bibnamefont {Heyl}},
  \bibinfo {author} {\bibfnamefont {M.}~\bibnamefont {Knap}}, \ and\ \bibinfo
  {author} {\bibfnamefont {A.}~\bibnamefont {Silva}},\ }\href {\doibase
  10.1103/PhysRevLett.120.130601} {\bibfield  {journal} {\bibinfo  {journal}
  {Phys. Rev. Lett.}\ }\textbf {\bibinfo {volume} {120}},\ \bibinfo {pages}
  {130601} (\bibinfo {year} {2018})}\BibitemShut {NoStop}%
\bibitem [{\citenamefont {Lerose}\ \emph {et~al.}(2018)\citenamefont {Lerose},
  \citenamefont {Marino}, \citenamefont {\ifmmode \check{Z}\else
  \v{Z}\fi{}unkovi\ifmmode~\check{c}\else \v{c}\fi{}}, \citenamefont
  {Gambassi},\ and\ \citenamefont {Silva}}]{Lerose2018Chaotic}%
  \BibitemOpen
  \bibfield  {author} {\bibinfo {author} {\bibfnamefont {A.}~\bibnamefont
  {Lerose}}, \bibinfo {author} {\bibfnamefont {J.}~\bibnamefont {Marino}},
  \bibinfo {author} {\bibfnamefont {B.}~\bibnamefont {\ifmmode \check{Z}\else
  \v{Z}\fi{}unkovi\ifmmode~\check{c}\else \v{c}\fi{}}}, \bibinfo {author}
  {\bibfnamefont {A.}~\bibnamefont {Gambassi}}, \ and\ \bibinfo {author}
  {\bibfnamefont {A.}~\bibnamefont {Silva}},\ }\href {\doibase
  10.1103/PhysRevLett.120.130603} {\bibfield  {journal} {\bibinfo  {journal}
  {Phys. Rev. Lett.}\ }\textbf {\bibinfo {volume} {120}},\ \bibinfo {pages}
  {130603} (\bibinfo {year} {2018})}\BibitemShut {NoStop}%
\bibitem [{\citenamefont {Lewis-Swan}\ \emph {et~al.}(2021)\citenamefont
  {Lewis-Swan}, \citenamefont {Muleady}, \citenamefont {Barberena},
  \citenamefont {Bollinger},\ and\ \citenamefont
  {Rey}}]{LewisSwan2021Characterizing}%
  \BibitemOpen
  \bibfield  {author} {\bibinfo {author} {\bibfnamefont {R.~J.}\ \bibnamefont
  {Lewis-Swan}}, \bibinfo {author} {\bibfnamefont {S.~R.}\ \bibnamefont
  {Muleady}}, \bibinfo {author} {\bibfnamefont {D.}~\bibnamefont {Barberena}},
  \bibinfo {author} {\bibfnamefont {J.~J.}\ \bibnamefont {Bollinger}}, \ and\
  \bibinfo {author} {\bibfnamefont {A.~M.}\ \bibnamefont {Rey}},\ }\href
  {\doibase 10.1103/PhysRevResearch.3.L022020} {\bibfield  {journal} {\bibinfo
  {journal} {Phys. Rev. Res.}\ }\textbf {\bibinfo {volume} {3}},\ \bibinfo
  {pages} {L022020} (\bibinfo {year} {2021})}\BibitemShut {NoStop}%
\bibitem [{\citenamefont {Parker}\ \emph {et~al.}(2019)\citenamefont {Parker},
  \citenamefont {Cao}, \citenamefont {Avdoshkin}, \citenamefont {Scaffidi},\
  and\ \citenamefont {Altman}}]{Parker2019Universal}%
  \BibitemOpen
  \bibfield  {author} {\bibinfo {author} {\bibfnamefont {D.~E.}\ \bibnamefont
  {Parker}}, \bibinfo {author} {\bibfnamefont {X.}~\bibnamefont {Cao}},
  \bibinfo {author} {\bibfnamefont {A.}~\bibnamefont {Avdoshkin}}, \bibinfo
  {author} {\bibfnamefont {T.}~\bibnamefont {Scaffidi}}, \ and\ \bibinfo
  {author} {\bibfnamefont {E.}~\bibnamefont {Altman}},\ }\href {\doibase
  10.1103/PhysRevX.9.041017} {\bibfield  {journal} {\bibinfo  {journal} {Phys.
  Rev. X}\ }\textbf {\bibinfo {volume} {9}},\ \bibinfo {pages} {041017}
  (\bibinfo {year} {2019})}\BibitemShut {NoStop}%
\bibitem [{\citenamefont {Balasubramanian}\ \emph {et~al.}(2022)\citenamefont
  {Balasubramanian}, \citenamefont {Caputa}, \citenamefont {Magan},\ and\
  \citenamefont {Wu}}]{Balasubramanian2022Quantum}%
  \BibitemOpen
  \bibfield  {author} {\bibinfo {author} {\bibfnamefont {V.}~\bibnamefont
  {Balasubramanian}}, \bibinfo {author} {\bibfnamefont {P.}~\bibnamefont
  {Caputa}}, \bibinfo {author} {\bibfnamefont {J.~M.}\ \bibnamefont {Magan}}, \
  and\ \bibinfo {author} {\bibfnamefont {Q.}~\bibnamefont {Wu}},\ }\href
  {\doibase 10.1103/PhysRevD.106.046007} {\bibfield  {journal} {\bibinfo
  {journal} {Phys. Rev. D}\ }\textbf {\bibinfo {volume} {106}},\ \bibinfo
  {pages} {046007} (\bibinfo {year} {2022})}\BibitemShut {NoStop}%
\bibitem [{\citenamefont {Rabinovici}\ \emph {et~al.}(2025)\citenamefont
  {Rabinovici}, \citenamefont {Sánchez-Garrido}, \citenamefont {Shir},\ and\
  \citenamefont {Sonner}}]{rabinovici2025krylovcomplexity}%
  \BibitemOpen
  \bibfield  {author} {\bibinfo {author} {\bibfnamefont {E.}~\bibnamefont
  {Rabinovici}}, \bibinfo {author} {\bibfnamefont {A.}~\bibnamefont
  {Sánchez-Garrido}}, \bibinfo {author} {\bibfnamefont {R.}~\bibnamefont
  {Shir}}, \ and\ \bibinfo {author} {\bibfnamefont {J.}~\bibnamefont
  {Sonner}},\ }\href {https://arxiv.org/abs/2507.06286} {\enquote {\bibinfo
  {title} {Krylov complexity},}\ } (\bibinfo {year} {2025}),\ \Eprint
  {http://arxiv.org/abs/2507.06286} {arXiv:2507.06286 [hep-th]} \BibitemShut
  {NoStop}%
\bibitem [{\citenamefont {Nandy}\ \emph {et~al.}(2025)\citenamefont {Nandy},
  \citenamefont {Matsoukas-Roubeas}, \citenamefont {Mart{\'\i}nez-Azcona},
  \citenamefont {Dymarsky},\ and\ \citenamefont {{del
  Campo}}}]{Nandy2025KrylovReview}%
  \BibitemOpen
  \bibfield  {author} {\bibinfo {author} {\bibfnamefont {P.}~\bibnamefont
  {Nandy}}, \bibinfo {author} {\bibfnamefont {A.~S.}\ \bibnamefont
  {Matsoukas-Roubeas}}, \bibinfo {author} {\bibfnamefont {P.}~\bibnamefont
  {Mart{\'\i}nez-Azcona}}, \bibinfo {author} {\bibfnamefont {A.}~\bibnamefont
  {Dymarsky}}, \ and\ \bibinfo {author} {\bibfnamefont {A.}~\bibnamefont {{del
  Campo}}},\ }\href {\doibase https://doi.org/10.1016/j.physrep.2025.05.001}
  {\bibfield  {journal} {\bibinfo  {journal} {Physics Reports}\ }\textbf
  {\bibinfo {volume} {1125-1128}},\ \bibinfo {pages} {1} (\bibinfo {year}
  {2025})},\ \bibinfo {note} {quantum dynamics in Krylov space: Methods and
  applications}\BibitemShut {NoStop}%
\bibitem [{\citenamefont {Craps}\ \emph {et~al.}(2024)\citenamefont {Craps},
  \citenamefont {Evnin},\ and\ \citenamefont {Pascuzzi}}]{Craps2024Relation}%
  \BibitemOpen
  \bibfield  {author} {\bibinfo {author} {\bibfnamefont {B.}~\bibnamefont
  {Craps}}, \bibinfo {author} {\bibfnamefont {O.}~\bibnamefont {Evnin}}, \ and\
  \bibinfo {author} {\bibfnamefont {G.}~\bibnamefont {Pascuzzi}},\ }\href
  {\doibase 10.1103/PhysRevLett.132.160402} {\bibfield  {journal} {\bibinfo
  {journal} {Phys. Rev. Lett.}\ }\textbf {\bibinfo {volume} {132}},\ \bibinfo
  {pages} {160402} (\bibinfo {year} {2024})}\BibitemShut {NoStop}%
\bibitem [{\citenamefont {Lv}\ \emph {et~al.}(2024)\citenamefont {Lv},
  \citenamefont {Zhang},\ and\ \citenamefont {Zhou}}]{Lv2024Building}%
  \BibitemOpen
  \bibfield  {author} {\bibinfo {author} {\bibfnamefont {C.}~\bibnamefont
  {Lv}}, \bibinfo {author} {\bibfnamefont {R.}~\bibnamefont {Zhang}}, \ and\
  \bibinfo {author} {\bibfnamefont {Q.}~\bibnamefont {Zhou}},\ }\href {\doibase
  10.1103/PhysRevResearch.6.L042001} {\bibfield  {journal} {\bibinfo  {journal}
  {Phys. Rev. Res.}\ }\textbf {\bibinfo {volume} {6}},\ \bibinfo {pages}
  {L042001} (\bibinfo {year} {2024})}\BibitemShut {NoStop}%
\bibitem [{\citenamefont {Rabinovici}\ \emph {et~al.}(2022)\citenamefont
  {Rabinovici}, \citenamefont {S{\'a}nchez-Garrido}, \citenamefont {Shir},\
  and\ \citenamefont {Sonner}}]{Rabinovici2022Krylov}%
  \BibitemOpen
  \bibfield  {author} {\bibinfo {author} {\bibfnamefont {E.}~\bibnamefont
  {Rabinovici}}, \bibinfo {author} {\bibfnamefont {A.}~\bibnamefont
  {S{\'a}nchez-Garrido}}, \bibinfo {author} {\bibfnamefont {R.}~\bibnamefont
  {Shir}}, \ and\ \bibinfo {author} {\bibfnamefont {J.}~\bibnamefont
  {Sonner}},\ }\href {\doibase 10.1007/JHEP07(2022)151} {\bibfield  {journal}
  {\bibinfo  {journal} {Journal of High Energy Physics}\ }\textbf {\bibinfo
  {volume} {2022}},\ \bibinfo {pages} {151} (\bibinfo {year}
  {2022})}\BibitemShut {NoStop}%
\bibitem [{\citenamefont {Espa\~nol}\ and\ \citenamefont
  {Wisniacki}(2023)}]{Espanol2023Assessing}%
  \BibitemOpen
  \bibfield  {author} {\bibinfo {author} {\bibfnamefont {B.~L.}\ \bibnamefont
  {Espa\~nol}}\ and\ \bibinfo {author} {\bibfnamefont {D.~A.}\ \bibnamefont
  {Wisniacki}},\ }\href {\doibase 10.1103/PhysRevE.107.024217} {\bibfield
  {journal} {\bibinfo  {journal} {Phys. Rev. E}\ }\textbf {\bibinfo {volume}
  {107}},\ \bibinfo {pages} {024217} (\bibinfo {year} {2023})}\BibitemShut
  {NoStop}%
\bibitem [{\citenamefont {Balasubramanian}\ \emph {et~al.}(2025)\citenamefont
  {Balasubramanian}, \citenamefont {Magan},\ and\ \citenamefont
  {Wu}}]{Balasubramanian2025Quantum}%
  \BibitemOpen
  \bibfield  {author} {\bibinfo {author} {\bibfnamefont {V.}~\bibnamefont
  {Balasubramanian}}, \bibinfo {author} {\bibfnamefont {J.~M.}\ \bibnamefont
  {Magan}}, \ and\ \bibinfo {author} {\bibfnamefont {Q.}~\bibnamefont {Wu}},\
  }\href {\doibase 10.1103/PhysRevE.111.014218} {\bibfield  {journal} {\bibinfo
   {journal} {Phys. Rev. E}\ }\textbf {\bibinfo {volume} {111}},\ \bibinfo
  {pages} {014218} (\bibinfo {year} {2025})}\BibitemShut {NoStop}%
\bibitem [{\citenamefont {Scialchi}\ \emph {et~al.}(2024)\citenamefont
  {Scialchi}, \citenamefont {Roncaglia},\ and\ \citenamefont
  {Wisniacki}}]{Scialchi2024Integrability}%
  \BibitemOpen
  \bibfield  {author} {\bibinfo {author} {\bibfnamefont {G.~F.}\ \bibnamefont
  {Scialchi}}, \bibinfo {author} {\bibfnamefont {A.~J.}\ \bibnamefont
  {Roncaglia}}, \ and\ \bibinfo {author} {\bibfnamefont {D.~A.}\ \bibnamefont
  {Wisniacki}},\ }\href {\doibase 10.1103/PhysRevE.109.054209} {\bibfield
  {journal} {\bibinfo  {journal} {Phys. Rev. E}\ }\textbf {\bibinfo {volume}
  {109}},\ \bibinfo {pages} {054209} (\bibinfo {year} {2024})}\BibitemShut
  {NoStop}%
\bibitem [{\citenamefont {Baggioli}\ \emph {et~al.}(2025)\citenamefont
  {Baggioli}, \citenamefont {Huh}, \citenamefont {Jeong}, \citenamefont {Kim},\
  and\ \citenamefont {Pedraza}}]{Baggioli2025Krylov}%
  \BibitemOpen
  \bibfield  {author} {\bibinfo {author} {\bibfnamefont {M.}~\bibnamefont
  {Baggioli}}, \bibinfo {author} {\bibfnamefont {K.-B.}\ \bibnamefont {Huh}},
  \bibinfo {author} {\bibfnamefont {H.-S.}\ \bibnamefont {Jeong}}, \bibinfo
  {author} {\bibfnamefont {K.-Y.}\ \bibnamefont {Kim}}, \ and\ \bibinfo
  {author} {\bibfnamefont {J.~F.}\ \bibnamefont {Pedraza}},\ }\href {\doibase
  10.1103/PhysRevResearch.7.023028} {\bibfield  {journal} {\bibinfo  {journal}
  {Phys. Rev. Res.}\ }\textbf {\bibinfo {volume} {7}},\ \bibinfo {pages}
  {023028} (\bibinfo {year} {2025})}\BibitemShut {NoStop}%
\bibitem [{\citenamefont {Camargo}\ \emph {et~al.}(2024)\citenamefont
  {Camargo}, \citenamefont {Huh}, \citenamefont {Jahnke}, \citenamefont
  {Jeong}, \citenamefont {Kim},\ and\ \citenamefont
  {Nishida}}]{Camargo2024Spread}%
  \BibitemOpen
  \bibfield  {author} {\bibinfo {author} {\bibfnamefont {H.~A.}\ \bibnamefont
  {Camargo}}, \bibinfo {author} {\bibfnamefont {K.-B.}\ \bibnamefont {Huh}},
  \bibinfo {author} {\bibfnamefont {V.}~\bibnamefont {Jahnke}}, \bibinfo
  {author} {\bibfnamefont {H.-S.}\ \bibnamefont {Jeong}}, \bibinfo {author}
  {\bibfnamefont {K.-Y.}\ \bibnamefont {Kim}}, \ and\ \bibinfo {author}
  {\bibfnamefont {M.}~\bibnamefont {Nishida}},\ }\href {\doibase
  10.1007/JHEP08(2024)241} {\bibfield  {journal} {\bibinfo  {journal} {Journal
  of High Energy Physics}\ }\textbf {\bibinfo {volume} {2024}},\ \bibinfo
  {pages} {241} (\bibinfo {year} {2024})}\BibitemShut {NoStop}%
\bibitem [{\citenamefont {Scialchi}\ \emph {et~al.}(2025)\citenamefont
  {Scialchi}, \citenamefont {Roncaglia}, \citenamefont {Pineda},\ and\
  \citenamefont {Wisniacki}}]{Scialchi2025Exploring}%
  \BibitemOpen
  \bibfield  {author} {\bibinfo {author} {\bibfnamefont {G.~F.}\ \bibnamefont
  {Scialchi}}, \bibinfo {author} {\bibfnamefont {A.~J.}\ \bibnamefont
  {Roncaglia}}, \bibinfo {author} {\bibfnamefont {C.}~\bibnamefont {Pineda}}, \
  and\ \bibinfo {author} {\bibfnamefont {D.~A.}\ \bibnamefont {Wisniacki}},\
  }\href {\doibase 10.1103/PhysRevE.111.014220} {\bibfield  {journal} {\bibinfo
   {journal} {Phys. Rev. E}\ }\textbf {\bibinfo {volume} {111}},\ \bibinfo
  {pages} {014220} (\bibinfo {year} {2025})}\BibitemShut {NoStop}%
\bibitem [{\citenamefont {Bento}\ \emph {et~al.}(2024)\citenamefont {Bento},
  \citenamefont {del Campo},\ and\ \citenamefont {C\'eleri}}]{Bento2024Krylov}%
  \BibitemOpen
  \bibfield  {author} {\bibinfo {author} {\bibfnamefont {P.~H.~S.}\
  \bibnamefont {Bento}}, \bibinfo {author} {\bibfnamefont {A.}~\bibnamefont
  {del Campo}}, \ and\ \bibinfo {author} {\bibfnamefont {L.~C.}\ \bibnamefont
  {C\'eleri}},\ }\href {\doibase 10.1103/PhysRevB.109.224304} {\bibfield
  {journal} {\bibinfo  {journal} {Phys. Rev. B}\ }\textbf {\bibinfo {volume}
  {109}},\ \bibinfo {pages} {224304} (\bibinfo {year} {2024})}\BibitemShut
  {NoStop}%
\bibitem [{\citenamefont {Dicke}(1954)}]{Dicke1954Coherence}%
  \BibitemOpen
  \bibfield  {author} {\bibinfo {author} {\bibfnamefont {R.~H.}\ \bibnamefont
  {Dicke}},\ }\href {\doibase 10.1103/PhysRev.93.99} {\bibfield  {journal}
  {\bibinfo  {journal} {Phys. Rev.}\ }\textbf {\bibinfo {volume} {93}},\
  \bibinfo {pages} {99} (\bibinfo {year} {1954})}\BibitemShut {NoStop}%
\bibitem [{\citenamefont {Safavi-Naini}\ \emph {et~al.}(2018)\citenamefont
  {Safavi-Naini}, \citenamefont {Lewis-Swan}, \citenamefont {Bohnet},
  \citenamefont {G\"arttner}, \citenamefont {Gilmore}, \citenamefont {Jordan},
  \citenamefont {Cohn}, \citenamefont {Freericks}, \citenamefont {Rey},\ and\
  \citenamefont {Bollinger}}]{Safavi2018Verification}%
  \BibitemOpen
  \bibfield  {author} {\bibinfo {author} {\bibfnamefont {A.}~\bibnamefont
  {Safavi-Naini}}, \bibinfo {author} {\bibfnamefont {R.~J.}\ \bibnamefont
  {Lewis-Swan}}, \bibinfo {author} {\bibfnamefont {J.~G.}\ \bibnamefont
  {Bohnet}}, \bibinfo {author} {\bibfnamefont {M.}~\bibnamefont {G\"arttner}},
  \bibinfo {author} {\bibfnamefont {K.~A.}\ \bibnamefont {Gilmore}}, \bibinfo
  {author} {\bibfnamefont {J.~E.}\ \bibnamefont {Jordan}}, \bibinfo {author}
  {\bibfnamefont {J.}~\bibnamefont {Cohn}}, \bibinfo {author} {\bibfnamefont
  {J.~K.}\ \bibnamefont {Freericks}}, \bibinfo {author} {\bibfnamefont {A.~M.}\
  \bibnamefont {Rey}}, \ and\ \bibinfo {author} {\bibfnamefont {J.~J.}\
  \bibnamefont {Bollinger}},\ }\href {\doibase 10.1103/PhysRevLett.121.040503}
  {\bibfield  {journal} {\bibinfo  {journal} {Phys. Rev. Lett.}\ }\textbf
  {\bibinfo {volume} {121}},\ \bibinfo {pages} {040503} (\bibinfo {year}
  {2018})}\BibitemShut {NoStop}%
\bibitem [{\citenamefont {Cohn}\ \emph {et~al.}(2018)\citenamefont {Cohn},
  \citenamefont {Safavi-Naini}, \citenamefont {Lewis-Swan}, \citenamefont
  {Bohnet}, \citenamefont {Gärttner}, \citenamefont {Gilmore}, \citenamefont
  {Jordan}, \citenamefont {Rey}, \citenamefont {Bollinger},\ and\ \citenamefont
  {Freericks}}]{Cohn2018Bang}%
  \BibitemOpen
  \bibfield  {author} {\bibinfo {author} {\bibfnamefont {J.}~\bibnamefont
  {Cohn}}, \bibinfo {author} {\bibfnamefont {A.}~\bibnamefont {Safavi-Naini}},
  \bibinfo {author} {\bibfnamefont {R.~J.}\ \bibnamefont {Lewis-Swan}},
  \bibinfo {author} {\bibfnamefont {J.~G.}\ \bibnamefont {Bohnet}}, \bibinfo
  {author} {\bibfnamefont {M.}~\bibnamefont {Gärttner}}, \bibinfo {author}
  {\bibfnamefont {K.~A.}\ \bibnamefont {Gilmore}}, \bibinfo {author}
  {\bibfnamefont {J.~E.}\ \bibnamefont {Jordan}}, \bibinfo {author}
  {\bibfnamefont {A.~M.}\ \bibnamefont {Rey}}, \bibinfo {author} {\bibfnamefont
  {J.~J.}\ \bibnamefont {Bollinger}}, \ and\ \bibinfo {author} {\bibfnamefont
  {J.~K.}\ \bibnamefont {Freericks}},\ }\href {\doibase
  10.1088/1367-2630/aac3fa} {\bibfield  {journal} {\bibinfo  {journal} {New
  Journal of Physics}\ }\textbf {\bibinfo {volume} {20}},\ \bibinfo {pages}
  {055013} (\bibinfo {year} {2018})}\BibitemShut {NoStop}%
\bibitem [{\citenamefont {Baumann}\ \emph {et~al.}(2010)\citenamefont
  {Baumann}, \citenamefont {Guerlin}, \citenamefont {Brennecke},\ and\
  \citenamefont {Esslinger}}]{Baumann2010Dicke}%
  \BibitemOpen
  \bibfield  {author} {\bibinfo {author} {\bibfnamefont {K.}~\bibnamefont
  {Baumann}}, \bibinfo {author} {\bibfnamefont {C.}~\bibnamefont {Guerlin}},
  \bibinfo {author} {\bibfnamefont {F.}~\bibnamefont {Brennecke}}, \ and\
  \bibinfo {author} {\bibfnamefont {T.}~\bibnamefont {Esslinger}},\ }\href
  {\doibase 10.1038/nature09009} {\bibfield  {journal} {\bibinfo  {journal}
  {Nature}\ }\textbf {\bibinfo {volume} {464}},\ \bibinfo {pages} {1301}
  (\bibinfo {year} {2010})}\BibitemShut {NoStop}%
\bibitem [{\citenamefont {Klinder}\ \emph {et~al.}(2015)\citenamefont
  {Klinder}, \citenamefont {Keßler}, \citenamefont {Wolke}, \citenamefont
  {Mathey},\ and\ \citenamefont {Hemmerich}}]{Klinder2015Dynamical}%
  \BibitemOpen
  \bibfield  {author} {\bibinfo {author} {\bibfnamefont {J.}~\bibnamefont
  {Klinder}}, \bibinfo {author} {\bibfnamefont {H.}~\bibnamefont {Keßler}},
  \bibinfo {author} {\bibfnamefont {M.}~\bibnamefont {Wolke}}, \bibinfo
  {author} {\bibfnamefont {L.}~\bibnamefont {Mathey}}, \ and\ \bibinfo {author}
  {\bibfnamefont {A.}~\bibnamefont {Hemmerich}},\ }\href {\doibase
  10.1073/pnas.1417132112} {\bibfield  {journal} {\bibinfo  {journal}
  {Proceedings of the National Academy of Sciences}\ }\textbf {\bibinfo
  {volume} {112}},\ \bibinfo {pages} {3290} (\bibinfo {year}
  {2015})}\BibitemShut {NoStop}%
\bibitem [{\citenamefont {Zhang}\ \emph {et~al.}(2018)\citenamefont {Zhang},
  \citenamefont {Lee}, \citenamefont {Kumar}, \citenamefont {Arnold},
  \citenamefont {Masson}, \citenamefont {Grimsmo}, \citenamefont {Parkins},\
  and\ \citenamefont {Barrett}}]{Zhang2018Dicke}%
  \BibitemOpen
  \bibfield  {author} {\bibinfo {author} {\bibfnamefont {Z.}~\bibnamefont
  {Zhang}}, \bibinfo {author} {\bibfnamefont {C.~H.}\ \bibnamefont {Lee}},
  \bibinfo {author} {\bibfnamefont {R.}~\bibnamefont {Kumar}}, \bibinfo
  {author} {\bibfnamefont {K.~J.}\ \bibnamefont {Arnold}}, \bibinfo {author}
  {\bibfnamefont {S.~J.}\ \bibnamefont {Masson}}, \bibinfo {author}
  {\bibfnamefont {A.~L.}\ \bibnamefont {Grimsmo}}, \bibinfo {author}
  {\bibfnamefont {A.~S.}\ \bibnamefont {Parkins}}, \ and\ \bibinfo {author}
  {\bibfnamefont {M.~D.}\ \bibnamefont {Barrett}},\ }\href {\doibase
  10.1103/PhysRevA.97.043858} {\bibfield  {journal} {\bibinfo  {journal} {Phys.
  Rev. A}\ }\textbf {\bibinfo {volume} {97}},\ \bibinfo {pages} {043858}
  (\bibinfo {year} {2018})}\BibitemShut {NoStop}%
\bibitem [{\citenamefont {Kroeze}\ \emph {et~al.}(2018)\citenamefont {Kroeze},
  \citenamefont {Guo}, \citenamefont {Vaidya}, \citenamefont {Keeling},\ and\
  \citenamefont {Lev}}]{Kroeze2018Spinor}%
  \BibitemOpen
  \bibfield  {author} {\bibinfo {author} {\bibfnamefont {R.~M.}\ \bibnamefont
  {Kroeze}}, \bibinfo {author} {\bibfnamefont {Y.}~\bibnamefont {Guo}},
  \bibinfo {author} {\bibfnamefont {V.~D.}\ \bibnamefont {Vaidya}}, \bibinfo
  {author} {\bibfnamefont {J.}~\bibnamefont {Keeling}}, \ and\ \bibinfo
  {author} {\bibfnamefont {B.~L.}\ \bibnamefont {Lev}},\ }\href {\doibase
  10.1103/PhysRevLett.121.163601} {\bibfield  {journal} {\bibinfo  {journal}
  {Phys. Rev. Lett.}\ }\textbf {\bibinfo {volume} {121}},\ \bibinfo {pages}
  {163601} (\bibinfo {year} {2018})}\BibitemShut {NoStop}%
\bibitem [{\citenamefont {Emary}\ and\ \citenamefont
  {Brandes}(2003{\natexlab{a}})}]{Emary2003Chaos}%
  \BibitemOpen
  \bibfield  {author} {\bibinfo {author} {\bibfnamefont {C.}~\bibnamefont
  {Emary}}\ and\ \bibinfo {author} {\bibfnamefont {T.}~\bibnamefont
  {Brandes}},\ }\href {\doibase 10.1103/PhysRevE.67.066203} {\bibfield
  {journal} {\bibinfo  {journal} {Phys. Rev. E}\ }\textbf {\bibinfo {volume}
  {67}},\ \bibinfo {pages} {066203} (\bibinfo {year}
  {2003}{\natexlab{a}})}\BibitemShut {NoStop}%
\bibitem [{\citenamefont {Emary}\ and\ \citenamefont
  {Brandes}(2003{\natexlab{b}})}]{Emary2023Quantum}%
  \BibitemOpen
  \bibfield  {author} {\bibinfo {author} {\bibfnamefont {C.}~\bibnamefont
  {Emary}}\ and\ \bibinfo {author} {\bibfnamefont {T.}~\bibnamefont
  {Brandes}},\ }\href {\doibase 10.1103/PhysRevLett.90.044101} {\bibfield
  {journal} {\bibinfo  {journal} {Phys. Rev. Lett.}\ }\textbf {\bibinfo
  {volume} {90}},\ \bibinfo {pages} {044101} (\bibinfo {year}
  {2003}{\natexlab{b}})}\BibitemShut {NoStop}%
\bibitem [{\citenamefont {Ch\'avez-Carlos}\ \emph
  {et~al.}(2016{\natexlab{a}})\citenamefont {Ch\'avez-Carlos}, \citenamefont
  {Bastarrachea-Magnani}, \citenamefont {Lerma-Hern\'andez},\ and\
  \citenamefont {Hirsch}}]{ChavezCarlos2016Classical}%
  \BibitemOpen
  \bibfield  {author} {\bibinfo {author} {\bibfnamefont {J.}~\bibnamefont
  {Ch\'avez-Carlos}}, \bibinfo {author} {\bibfnamefont {M.~A.}\ \bibnamefont
  {Bastarrachea-Magnani}}, \bibinfo {author} {\bibfnamefont {S.}~\bibnamefont
  {Lerma-Hern\'andez}}, \ and\ \bibinfo {author} {\bibfnamefont {J.~G.}\
  \bibnamefont {Hirsch}},\ }\href {\doibase 10.1103/PhysRevE.94.022209}
  {\bibfield  {journal} {\bibinfo  {journal} {Phys. Rev. E}\ }\textbf {\bibinfo
  {volume} {94}},\ \bibinfo {pages} {022209} (\bibinfo {year}
  {2016}{\natexlab{a}})}\BibitemShut {NoStop}%
\bibitem [{\citenamefont {Lewis-Swan}\ \emph {et~al.}(2019)\citenamefont
  {Lewis-Swan}, \citenamefont {Safavi-Naini}, \citenamefont {Bollinger},\ and\
  \citenamefont {Rey}}]{LewisSwan2019Scrambling}%
  \BibitemOpen
  \bibfield  {author} {\bibinfo {author} {\bibfnamefont {R.~J.}\ \bibnamefont
  {Lewis-Swan}}, \bibinfo {author} {\bibfnamefont {A.}~\bibnamefont
  {Safavi-Naini}}, \bibinfo {author} {\bibfnamefont {J.~J.}\ \bibnamefont
  {Bollinger}}, \ and\ \bibinfo {author} {\bibfnamefont {A.~M.}\ \bibnamefont
  {Rey}},\ }\href {\doibase 10.1038/s41467-019-09436-y} {\bibfield  {journal}
  {\bibinfo  {journal} {Nature Communications}\ }\textbf {\bibinfo {volume}
  {10}},\ \bibinfo {pages} {1581} (\bibinfo {year} {2019})}\BibitemShut
  {NoStop}%
\bibitem [{\citenamefont {Pilatowsky-Cameo}\ \emph {et~al.}(2020)\citenamefont
  {Pilatowsky-Cameo}, \citenamefont {Ch\'avez-Carlos}, \citenamefont
  {Bastarrachea-Magnani}, \citenamefont {Str\'ansk\'y}, \citenamefont
  {Lerma-Hern\'andez}, \citenamefont {Santos},\ and\ \citenamefont
  {Hirsch}}]{PilatowskyCameo2020Positive}%
  \BibitemOpen
  \bibfield  {author} {\bibinfo {author} {\bibfnamefont {S.}~\bibnamefont
  {Pilatowsky-Cameo}}, \bibinfo {author} {\bibfnamefont {J.}~\bibnamefont
  {Ch\'avez-Carlos}}, \bibinfo {author} {\bibfnamefont {M.~A.}\ \bibnamefont
  {Bastarrachea-Magnani}}, \bibinfo {author} {\bibfnamefont {P.}~\bibnamefont
  {Str\'ansk\'y}}, \bibinfo {author} {\bibfnamefont {S.}~\bibnamefont
  {Lerma-Hern\'andez}}, \bibinfo {author} {\bibfnamefont {L.~F.}\ \bibnamefont
  {Santos}}, \ and\ \bibinfo {author} {\bibfnamefont {J.~G.}\ \bibnamefont
  {Hirsch}},\ }\href {\doibase 10.1103/PhysRevE.101.010202} {\bibfield
  {journal} {\bibinfo  {journal} {Phys. Rev. E}\ }\textbf {\bibinfo {volume}
  {101}},\ \bibinfo {pages} {010202(R)} (\bibinfo {year} {2020})}\BibitemShut
  {NoStop}%
\bibitem [{\citenamefont {Villase{\~n}or}\ \emph {et~al.}(2023)\citenamefont
  {Villase{\~n}or}, \citenamefont {Pilatowsky-Cameo}, \citenamefont
  {Bastarrachea-Magnani}, \citenamefont {Lerma-Hern{\'a}ndez}, \citenamefont
  {Santos},\ and\ \citenamefont {Hirsch}}]{Villasenor2023Chaos}%
  \BibitemOpen
  \bibfield  {author} {\bibinfo {author} {\bibfnamefont {D.}~\bibnamefont
  {Villase{\~n}or}}, \bibinfo {author} {\bibfnamefont {S.}~\bibnamefont
  {Pilatowsky-Cameo}}, \bibinfo {author} {\bibfnamefont {M.~A.}\ \bibnamefont
  {Bastarrachea-Magnani}}, \bibinfo {author} {\bibfnamefont {S.}~\bibnamefont
  {Lerma-Hern{\'a}ndez}}, \bibinfo {author} {\bibfnamefont {L.~F.}\
  \bibnamefont {Santos}}, \ and\ \bibinfo {author} {\bibfnamefont {J.~G.}\
  \bibnamefont {Hirsch}},\ }\href {\doibase 10.3390/e25010008} {\bibfield
  {journal} {\bibinfo  {journal} {Entropy}\ }\textbf {\bibinfo {volume} {25}}
  (\bibinfo {year} {2023}),\ 10.3390/e25010008}\BibitemShut {NoStop}%
\bibitem [{\citenamefont {Villaseñor}\ \emph {et~al.}(2026)\citenamefont
  {Villaseñor}, \citenamefont {Pilatowsky-Cameo}, \citenamefont
  {Chávez-Carlos}, \citenamefont {Bastarrachea-Magnani}, \citenamefont
  {Lerma-Hernández}, \citenamefont {Santos},\ and\ \citenamefont
  {Hirsch}}]{villasenor2026classical}%
  \BibitemOpen
  \bibfield  {author} {\bibinfo {author} {\bibfnamefont {D.}~\bibnamefont
  {Villaseñor}}, \bibinfo {author} {\bibfnamefont {S.}~\bibnamefont
  {Pilatowsky-Cameo}}, \bibinfo {author} {\bibfnamefont {J.}~\bibnamefont
  {Chávez-Carlos}}, \bibinfo {author} {\bibfnamefont {M.~A.}\ \bibnamefont
  {Bastarrachea-Magnani}}, \bibinfo {author} {\bibfnamefont {S.}~\bibnamefont
  {Lerma-Hernández}}, \bibinfo {author} {\bibfnamefont {L.~F.}\ \bibnamefont
  {Santos}}, \ and\ \bibinfo {author} {\bibfnamefont {J.~G.}\ \bibnamefont
  {Hirsch}},\ }\href {https://arxiv.org/abs/2405.20381} {\enquote {\bibinfo
  {title} {Classical and quantum properties of the spin-boson dicke model:
  Chaos, localization, and scarring},}\ } (\bibinfo {year} {2026}),\ \Eprint
  {http://arxiv.org/abs/2405.20381} {arXiv:2405.20381 [quant-ph]} \BibitemShut
  {NoStop}%
\bibitem [{\citenamefont {Tavis}\ and\ \citenamefont
  {Cummings}(1968)}]{Tavis1968Exact}%
  \BibitemOpen
  \bibfield  {author} {\bibinfo {author} {\bibfnamefont {M.}~\bibnamefont
  {Tavis}}\ and\ \bibinfo {author} {\bibfnamefont {F.~W.}\ \bibnamefont
  {Cummings}},\ }\href {\doibase 10.1103/PhysRev.170.379} {\bibfield  {journal}
  {\bibinfo  {journal} {Phys. Rev.}\ }\textbf {\bibinfo {volume} {170}},\
  \bibinfo {pages} {379} (\bibinfo {year} {1968})}\BibitemShut {NoStop}%
\bibitem [{\citenamefont {Lipkin}\ \emph {et~al.}(1965)\citenamefont {Lipkin},
  \citenamefont {Meshkov},\ and\ \citenamefont {Glick}}]{LMG1965}%
  \BibitemOpen
  \bibfield  {author} {\bibinfo {author} {\bibfnamefont {H.}~\bibnamefont
  {Lipkin}}, \bibinfo {author} {\bibfnamefont {N.}~\bibnamefont {Meshkov}}, \
  and\ \bibinfo {author} {\bibfnamefont {A.}~\bibnamefont {Glick}},\ }\href
  {\doibase https://doi.org/10.1016/0029-5582(65)90862-X} {\bibfield  {journal}
  {\bibinfo  {journal} {Nuclear Physics}\ }\textbf {\bibinfo {volume} {62}},\
  \bibinfo {pages} {188} (\bibinfo {year} {1965})}\BibitemShut {NoStop}%
\bibitem [{\citenamefont {Safavi-Naini}\ \emph {et~al.}(2017)\citenamefont
  {Safavi-Naini}, \citenamefont {Lewis-Swan}, \citenamefont {Bohnet},
  \citenamefont {Garttner}, \citenamefont {Gilmore}, \citenamefont {Jordan},
  \citenamefont {Cohn}, \citenamefont {Freericks}, \citenamefont {Rey},\ and\
  \citenamefont {Bollinger}}]{SafaviNaini2017dicke}%
  \BibitemOpen
  \bibfield  {author} {\bibinfo {author} {\bibfnamefont {A.}~\bibnamefont
  {Safavi-Naini}}, \bibinfo {author} {\bibfnamefont {R.~J.}\ \bibnamefont
  {Lewis-Swan}}, \bibinfo {author} {\bibfnamefont {J.~G.}\ \bibnamefont
  {Bohnet}}, \bibinfo {author} {\bibfnamefont {M.}~\bibnamefont {Garttner}},
  \bibinfo {author} {\bibfnamefont {K.~A.}\ \bibnamefont {Gilmore}}, \bibinfo
  {author} {\bibfnamefont {E.}~\bibnamefont {Jordan}}, \bibinfo {author}
  {\bibfnamefont {J.}~\bibnamefont {Cohn}}, \bibinfo {author} {\bibfnamefont
  {J.~K.}\ \bibnamefont {Freericks}}, \bibinfo {author} {\bibfnamefont {A.~M.}\
  \bibnamefont {Rey}}, \ and\ \bibinfo {author} {\bibfnamefont {J.~J.}\
  \bibnamefont {Bollinger}},\ }\href@noop {} {\  (\bibinfo {year} {2017})},\
  \Eprint {http://arxiv.org/abs/1711.07392} {arXiv:1711.07392 [quant-ph]}
  \BibitemShut {NoStop}%
\bibitem [{\citenamefont {Gilmore}\ \emph {et~al.}(2021)\citenamefont
  {Gilmore}, \citenamefont {Affolter}, \citenamefont {Lewis-Swan},
  \citenamefont {Barberena}, \citenamefont {Jordan}, \citenamefont {Rey},\ and\
  \citenamefont {Bollinger}}]{Gilmore2021Harnessing}%
  \BibitemOpen
  \bibfield  {author} {\bibinfo {author} {\bibfnamefont {K.~A.}\ \bibnamefont
  {Gilmore}}, \bibinfo {author} {\bibfnamefont {M.}~\bibnamefont {Affolter}},
  \bibinfo {author} {\bibfnamefont {R.~J.}\ \bibnamefont {Lewis-Swan}},
  \bibinfo {author} {\bibfnamefont {D.}~\bibnamefont {Barberena}}, \bibinfo
  {author} {\bibfnamefont {E.}~\bibnamefont {Jordan}}, \bibinfo {author}
  {\bibfnamefont {A.~M.}\ \bibnamefont {Rey}}, \ and\ \bibinfo {author}
  {\bibfnamefont {J.~J.}\ \bibnamefont {Bollinger}},\ }\href {\doibase
  10.1126/science.abi5226} {\bibfield  {journal} {\bibinfo  {journal}
  {Science}\ }\textbf {\bibinfo {volume} {373}},\ \bibinfo {pages} {673}
  (\bibinfo {year} {2021})}\BibitemShut {NoStop}%
\bibitem [{\citenamefont {Zhang}\ \emph {et~al.}(2025)\citenamefont {Zhang},
  \citenamefont {Zu\~niga Castro},\ and\ \citenamefont
  {Lewis-Swan}}]{Zhang2025Harnessing}%
  \BibitemOpen
  \bibfield  {author} {\bibinfo {author} {\bibfnamefont {Y.}~\bibnamefont
  {Zhang}}, \bibinfo {author} {\bibfnamefont {J.~C.}\ \bibnamefont {Zu\~niga
  Castro}}, \ and\ \bibinfo {author} {\bibfnamefont {R.~J.}\ \bibnamefont
  {Lewis-Swan}},\ }\href {\doibase 10.1103/PhysRevResearch.7.013227} {\bibfield
   {journal} {\bibinfo  {journal} {Phys. Rev. Res.}\ }\textbf {\bibinfo
  {volume} {7}},\ \bibinfo {pages} {013227} (\bibinfo {year}
  {2025})}\BibitemShut {NoStop}%
\bibitem [{\citenamefont {Lanczos}(1950)}]{Lanczos1950}%
  \BibitemOpen
  \bibfield  {author} {\bibinfo {author} {\bibfnamefont {C.}~\bibnamefont
  {Lanczos}},\ }\href {\doibase 10.6028/jres.045.026} {\bibfield  {journal}
  {\bibinfo  {journal} {Journal of Research of the National Bureau of
  Standards}\ }\textbf {\bibinfo {volume} {45}},\ \bibinfo {pages} {255}
  (\bibinfo {year} {1950})}\BibitemShut {NoStop}%
\bibitem [{\citenamefont {Viswanath}\ and\ \citenamefont
  {Mueller}(1994)}]{viswanath_mueller_1994}%
  \BibitemOpen
  \bibfield  {author} {\bibinfo {author} {\bibfnamefont {V.}~\bibnamefont
  {Viswanath}}\ and\ \bibinfo {author} {\bibfnamefont {G.}~\bibnamefont
  {Mueller}},\ }\href@noop {} {\emph {\bibinfo {title} {The recursion method.
  Application to many-body dynamics}}},\ Vol.~\bibinfo {volume} {23}\ (\bibinfo
   {publisher} {Springer.},\ \bibinfo {year} {1994})\BibitemShut {NoStop}%
\bibitem [{SM()}]{SM}%
  \BibitemOpen
  \href@noop {} {\enquote {\bibinfo {title} {{See Supplemental Material at [URL
  will be inserted by publisher].}}}\ }\BibitemShut {NoStop}%
\bibitem [{\citenamefont {Ch\'avez-Carlos}\ \emph
  {et~al.}(2016{\natexlab{b}})\citenamefont {Ch\'avez-Carlos}, \citenamefont
  {Bastarrachea-Magnani}, \citenamefont {Lerma-Hern\'andez},\ and\
  \citenamefont {Hirsch}}]{Chavez2016Classical}%
  \BibitemOpen
  \bibfield  {author} {\bibinfo {author} {\bibfnamefont {J.}~\bibnamefont
  {Ch\'avez-Carlos}}, \bibinfo {author} {\bibfnamefont {M.~A.}\ \bibnamefont
  {Bastarrachea-Magnani}}, \bibinfo {author} {\bibfnamefont {S.}~\bibnamefont
  {Lerma-Hern\'andez}}, \ and\ \bibinfo {author} {\bibfnamefont {J.~G.}\
  \bibnamefont {Hirsch}},\ }\href {\doibase 10.1103/PhysRevE.94.022209}
  {\bibfield  {journal} {\bibinfo  {journal} {Phys. Rev. E}\ }\textbf {\bibinfo
  {volume} {94}},\ \bibinfo {pages} {022209} (\bibinfo {year}
  {2016}{\natexlab{b}})}\BibitemShut {NoStop}%
\bibitem [{\citenamefont {Skokos}(2010)}]{Skokos2010Book}%
  \BibitemOpen
  \bibfield  {author} {\bibinfo {author} {\bibfnamefont {C.}~\bibnamefont
  {Skokos}},\ }\enquote {\bibinfo {title} {The {L}yapunov characteristic
  exponents and their computation},}\ in\ \href {\doibase
  10.1007/978-3-642-04458-8_2} {\emph {\bibinfo {booktitle} {Dynamics of Small
  Solar System Bodies and Exoplanets}}},\ \bibinfo {editor} {edited by\
  \bibinfo {editor} {\bibfnamefont {J.~J.}\ \bibnamefont {Souchay}}\ and\
  \bibinfo {editor} {\bibfnamefont {R.}~\bibnamefont {Dvorak}}}\ (\bibinfo
  {publisher} {Springer Berlin Heidelberg},\ \bibinfo {address} {Berlin,
  Heidelberg},\ \bibinfo {year} {2010})\ pp.\ \bibinfo {pages}
  {63--135}\BibitemShut {NoStop}%
\bibitem [{\citenamefont {Rodr{\'\i}guez-Laguna}\ \emph
  {et~al.}(2017)\citenamefont {Rodr{\'\i}guez-Laguna}, \citenamefont
  {Tarruell}, \citenamefont {Lewenstein}, \citenamefont {Asaduzzaman},\ and\
  \citenamefont {Celi}}]{rodriguez2017synthetic}%
  \BibitemOpen
  \bibfield  {author} {\bibinfo {author} {\bibfnamefont {J.}~\bibnamefont
  {Rodr{\'\i}guez-Laguna}}, \bibinfo {author} {\bibfnamefont {L.}~\bibnamefont
  {Tarruell}}, \bibinfo {author} {\bibfnamefont {M.}~\bibnamefont
  {Lewenstein}}, \bibinfo {author} {\bibfnamefont {M.}~\bibnamefont
  {Asaduzzaman}}, \ and\ \bibinfo {author} {\bibfnamefont {A.}~\bibnamefont
  {Celi}},\ }\href {\doibase 10.1103/PhysRevA.95.013627} {\bibfield  {journal}
  {\bibinfo  {journal} {Physical Review A}\ }\textbf {\bibinfo {volume} {95}},\
  \bibinfo {pages} {013627} (\bibinfo {year} {2017})}\BibitemShut {NoStop}%
\bibitem [{\citenamefont {Morice}\ \emph {et~al.}(2021)\citenamefont {Morice},
  \citenamefont {Chernyavsky}, \citenamefont {Moghaddam}, \citenamefont
  {van~den Brink},\ and\ \citenamefont {van Wezel}}]{morice2021synthetic}%
  \BibitemOpen
  \bibfield  {author} {\bibinfo {author} {\bibfnamefont {C.}~\bibnamefont
  {Morice}}, \bibinfo {author} {\bibfnamefont {D.}~\bibnamefont {Chernyavsky}},
  \bibinfo {author} {\bibfnamefont {A.~G.}\ \bibnamefont {Moghaddam}}, \bibinfo
  {author} {\bibfnamefont {J.}~\bibnamefont {van~den Brink}}, \ and\ \bibinfo
  {author} {\bibfnamefont {J.}~\bibnamefont {van Wezel}},\ }\href {\doibase
  10.1103/PhysRevResearch.3.L022022} {\bibfield  {journal} {\bibinfo  {journal}
  {Physical Review Research}\ }\textbf {\bibinfo {volume} {3}},\ \bibinfo
  {pages} {L022022} (\bibinfo {year} {2021})}\BibitemShut {NoStop}%
\bibitem [{\citenamefont {Morice}\ \emph {et~al.}(2022)\citenamefont {Morice},
  \citenamefont {Chernyavsky}, \citenamefont {Moghaddam}, \citenamefont
  {van~den Brink},\ and\ \citenamefont {van Wezel}}]{morice2022quantum}%
  \BibitemOpen
  \bibfield  {author} {\bibinfo {author} {\bibfnamefont {C.}~\bibnamefont
  {Morice}}, \bibinfo {author} {\bibfnamefont {D.}~\bibnamefont {Chernyavsky}},
  \bibinfo {author} {\bibfnamefont {A.~G.}\ \bibnamefont {Moghaddam}}, \bibinfo
  {author} {\bibfnamefont {J.}~\bibnamefont {van~den Brink}}, \ and\ \bibinfo
  {author} {\bibfnamefont {J.}~\bibnamefont {van Wezel}},\ }\href {\doibase
  10.21468/SciPostPhysCore.5.3.042} {\bibfield  {journal} {\bibinfo  {journal}
  {SciPost Physics Core}\ }\textbf {\bibinfo {volume} {5}},\ \bibinfo {pages}
  {042} (\bibinfo {year} {2022})}\BibitemShut {NoStop}%
\bibitem [{\citenamefont {Lv}\ \emph {et~al.}(2022)\citenamefont {Lv},
  \citenamefont {Zhang}, \citenamefont {Zhai},\ and\ \citenamefont
  {Zhou}}]{lv2022curving}%
  \BibitemOpen
  \bibfield  {author} {\bibinfo {author} {\bibfnamefont {C.}~\bibnamefont
  {Lv}}, \bibinfo {author} {\bibfnamefont {R.}~\bibnamefont {Zhang}}, \bibinfo
  {author} {\bibfnamefont {Z.}~\bibnamefont {Zhai}}, \ and\ \bibinfo {author}
  {\bibfnamefont {Q.}~\bibnamefont {Zhou}},\ }\href@noop {} {\bibfield
  {journal} {\bibinfo  {journal} {Nature communications}\ }\textbf {\bibinfo
  {volume} {13}},\ \bibinfo {pages} {2184} (\bibinfo {year}
  {2022})}\BibitemShut {NoStop}%
\bibitem [{\citenamefont {Tallent}\ and\ \citenamefont
  {Sheehy}(2025)}]{analog-unruh}%
  \BibitemOpen
  \bibfield  {author} {\bibinfo {author} {\bibfnamefont {K.~B.}\ \bibnamefont
  {Tallent}}\ and\ \bibinfo {author} {\bibfnamefont {D.~E.}\ \bibnamefont
  {Sheehy}},\ }\href {\doibase 10.1103/yccb-db4j} {\bibfield  {journal}
  {\bibinfo  {journal} {Phys. Rev. B}\ }\textbf {\bibinfo {volume} {112}},\
  \bibinfo {pages} {104306} (\bibinfo {year} {2025})}\BibitemShut {NoStop}%
\bibitem [{\citenamefont {Barcelo}\ \emph {et~al.}(2011)\citenamefont
  {Barcelo}, \citenamefont {Liberati},\ and\ \citenamefont
  {Visser}}]{barcelo2011analogue}%
  \BibitemOpen
  \bibfield  {author} {\bibinfo {author} {\bibfnamefont {C.}~\bibnamefont
  {Barcelo}}, \bibinfo {author} {\bibfnamefont {S.}~\bibnamefont {Liberati}}, \
  and\ \bibinfo {author} {\bibfnamefont {M.}~\bibnamefont {Visser}},\
  }\href@noop {} {\bibfield  {journal} {\bibinfo  {journal} {Living Reviews in
  Relativity}\ }\textbf {\bibinfo {volume} {14}},\ \bibinfo {pages} {3}
  (\bibinfo {year} {2011})}\BibitemShut {NoStop}%
\bibitem [{\citenamefont {Rindler}(1960)}]{rindler}%
  \BibitemOpen
  \bibfield  {author} {\bibinfo {author} {\bibfnamefont {W.}~\bibnamefont
  {Rindler}},\ }\href {\doibase 10.1103/PhysRev.119.2082} {\bibfield  {journal}
  {\bibinfo  {journal} {Phys. Rev.}\ }\textbf {\bibinfo {volume} {119}},\
  \bibinfo {pages} {2082} (\bibinfo {year} {1960})}\BibitemShut {NoStop}%
\bibitem [{\citenamefont {Misner}\ \emph {et~al.}(1973)\citenamefont {Misner},
  \citenamefont {Thorne},\ and\ \citenamefont
  {Wheeler}}]{misner1973gravitation}%
  \BibitemOpen
  \bibfield  {author} {\bibinfo {author} {\bibfnamefont {C.~W.}\ \bibnamefont
  {Misner}}, \bibinfo {author} {\bibfnamefont {K.~S.}\ \bibnamefont {Thorne}},
  \ and\ \bibinfo {author} {\bibfnamefont {J.~A.}\ \bibnamefont {Wheeler}},\
  }\href@noop {} {\emph {\bibinfo {title} {Gravitation}}}\ (\bibinfo
  {publisher} {Macmillan},\ \bibinfo {year} {1973})\BibitemShut {NoStop}%
\bibitem [{\citenamefont {Hawking}(2009)}]{hawking2009brief}%
  \BibitemOpen
  \bibfield  {author} {\bibinfo {author} {\bibfnamefont {S.}~\bibnamefont
  {Hawking}},\ }\href@noop {} {\emph {\bibinfo {title} {A brief history of
  time: from big bang to black holes}}}\ (\bibinfo  {publisher} {Random
  House},\ \bibinfo {year} {2009})\BibitemShut {NoStop}%
\bibitem [{\citenamefont {Pinochet}(2022)}]{pinochet2022little}%
  \BibitemOpen
  \bibfield  {author} {\bibinfo {author} {\bibfnamefont {J.}~\bibnamefont
  {Pinochet}},\ }\href@noop {} {\bibfield  {journal} {\bibinfo  {journal}
  {Physics Education}\ }\textbf {\bibinfo {volume} {57}},\ \bibinfo {pages}
  {045008} (\bibinfo {year} {2022})}\BibitemShut {NoStop}%
\bibitem [{Note1()}]{Note1}%
  \BibitemOpen
  \bibinfo {note} {In Ref.~\cite {Balasubramanian2022Quantum}, the transition
  is at $\omega =0$ from the standard to the inverted oscillator, which are
  connected by analytical continuation $\omega \rightarrow -i\omega $. The
  model and motivation are different from our case.}\BibitemShut {Stop}%
\bibitem [{\citenamefont {Oganesyan}\ and\ \citenamefont
  {Huse}(2007)}]{Oganesyan2007Localization}%
  \BibitemOpen
  \bibfield  {author} {\bibinfo {author} {\bibfnamefont {V.}~\bibnamefont
  {Oganesyan}}\ and\ \bibinfo {author} {\bibfnamefont {D.~A.}\ \bibnamefont
  {Huse}},\ }\href {\doibase 10.1103/PhysRevB.75.155111} {\bibfield  {journal}
  {\bibinfo  {journal} {Phys. Rev. B}\ }\textbf {\bibinfo {volume} {75}},\
  \bibinfo {pages} {155111} (\bibinfo {year} {2007})}\BibitemShut {NoStop}%
\bibitem [{\citenamefont {Moghaddam}\ \emph {et~al.}(2021)\citenamefont
  {Moghaddam}, \citenamefont {Chernyavsky}, \citenamefont {Morice},
  \citenamefont {van Wezel},\ and\ \citenamefont {van~den
  Brink}}]{moghaddam2021engineering}%
  \BibitemOpen
  \bibfield  {author} {\bibinfo {author} {\bibfnamefont {A.~G.}\ \bibnamefont
  {Moghaddam}}, \bibinfo {author} {\bibfnamefont {D.}~\bibnamefont
  {Chernyavsky}}, \bibinfo {author} {\bibfnamefont {C.}~\bibnamefont {Morice}},
  \bibinfo {author} {\bibfnamefont {J.}~\bibnamefont {van Wezel}}, \ and\
  \bibinfo {author} {\bibfnamefont {J.}~\bibnamefont {van~den Brink}},\ }\href
  {\doibase 10.21468/SciPostPhys.11.6.109} {\bibfield  {journal} {\bibinfo
  {journal} {SciPost Physics}\ }\textbf {\bibinfo {volume} {11}},\ \bibinfo
  {pages} {109} (\bibinfo {year} {2021})}\BibitemShut {NoStop}%
\end{thebibliography}%

\clearpage
\appendix
\section*{End Matter}
\begin{figure}[!bt]
\includegraphics[width=1\columnwidth]{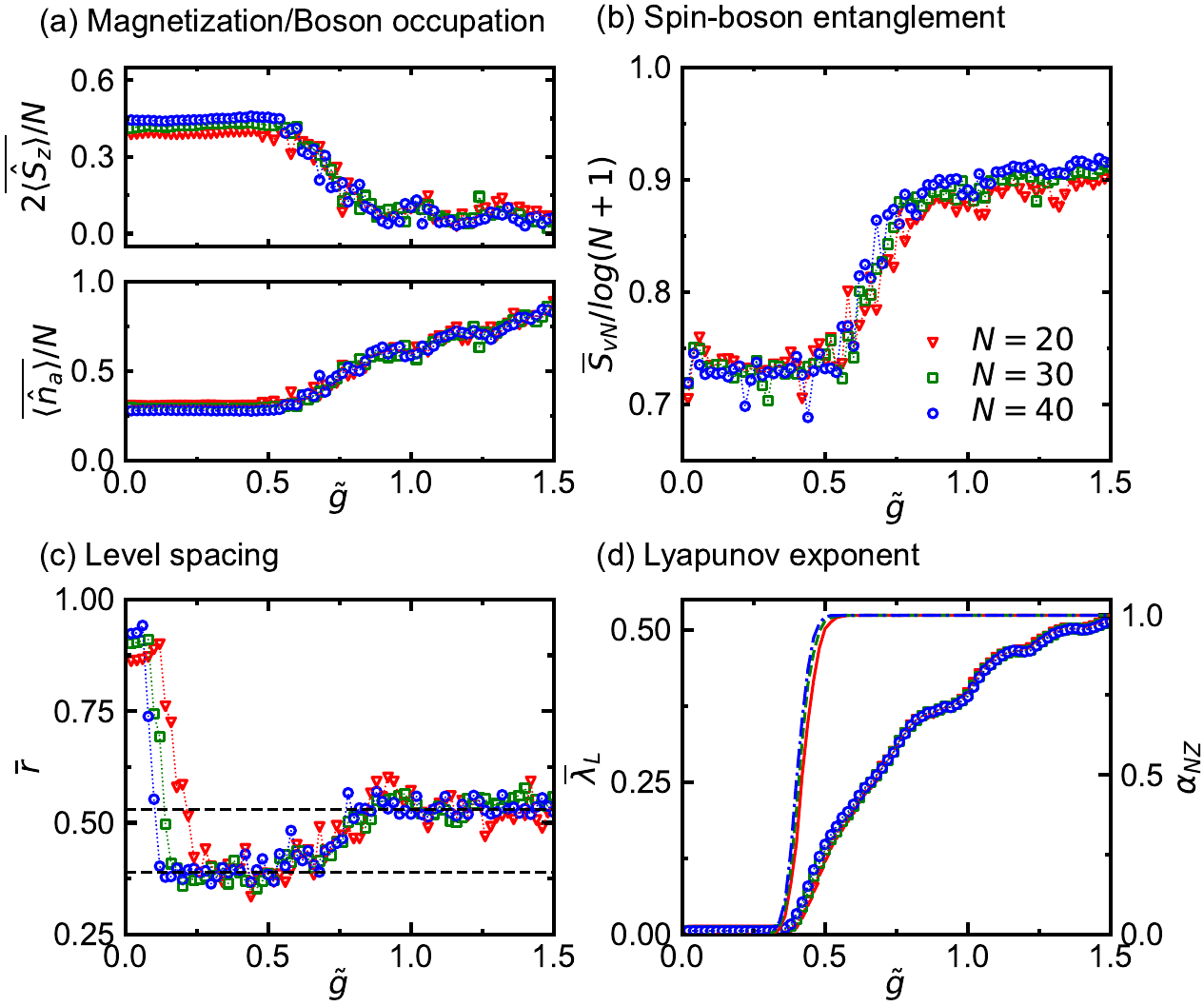}
    \caption{(a) The average of magnetization $\hat S_z$ and boson occupation $\hat n_a$ over time interval $gt\in[50,200]$. (b) Time-averaged spin-boson entanglement entropy ${S}_{\rm vN}$. (c) The mean ratio of consecutive level spacings $\overline r$ within a finite window containing the initial energy. Horizontal dashed lines are $\overline r=0.386$ (Poisson) and $0.536$ (Wigner-Dyson). (d) Symbols: Lyapunov exponent ${\lambda}_L$ averaged over sampled initial states. Lines: $\alpha_{\rm NZ}$, the fraction of initial states with finite $\lambda_L>0.01$. $N=20$ (triangles), 30 (squares), and 40 (circles); $\delta=\Omega$.}
    \label{fig:transition}
\end{figure}

{\it Onset of chaos}---The dichotomy between integrable and chaotic dynamics is often blurred and bridged by 
a smooth crossover. This stands in contrast to a DPT, where certain observables undergo sharp changes. Fig. \ref{fig:transition} summarizes data from traditional probes of DPT or chaos. They serve two purposes: (1) to show broad consistency with the $C_K$ and $S_K$ results in the main text; (2) to illustrate the advantages offered by Krylov complexity. Fig. \ref{fig:transition}(a) shows the time average of $S_z$ and $n_b$. Note that the flat plateau in the regular regime is a unique feature of being at resonance $\Omega=\delta$, and both averages show pronounced fluctuations at larger $\tilde{g}$. Fig.~\ref{fig:transition}(b) plots the time average of spin-boson entanglement entropy $S_{vN}=-{\rm Tr}\rho_{\rm s}\ln\rho_{\rm s}$, where $\rho_{\rm s}$ is the reduced density matrix for the spin degrees of freedom \cite{SM}. The regular regime features lower entanglement; in fact $S_{vN}(t)$ oscillates just like other observables so the system can return to low-entropy states where the spin and boson degrees of freedom are separable. For $g>\tilde g_c$, $\overline S_{vN}$ rapidly rises to $\sim90\%$ of the maximum entanglement $\ln(N+1)$.

To probe the onset of chaos, we use a standard tool called the mean ratio of consecutive level spacings. Define $r_i={\rm min}(\delta^E_{i},\, \delta^E_{i-1})/{\rm max}(\delta^E_{i},\, \delta^E_{i-1})$, where $\delta^E_{i}=E_{i}-E_{i-1}$ is the spacing between two consecutive energy levels \cite{Oganesyan2007Localization}. Typically, one averages $r_i$ over the entire spectrum, which is obviously unfit for our purpose. Instead, we restrict the average $\overline r$ to a ``microcanonical ensemble" of $6N$ states near $E_0$, the energy of the initial state. As seen in Fig.~\ref{fig:transition}(c), except for very small $\tilde g$, where $\overline r$ is close to 1 because the energy levels are almost equally spaced, we find $\overline r\sim0.386$ up until $\tilde g_c$, i.e., the level statistics obeys the Poisson distribution. For $\tilde{g}>\tilde g_c$, $\overline r\sim0.536$, which is consistent with the Wigner-Dyson distribution of chaotic Hamiltonians. This further justifies calling the latter the ``chaotic" regime.

To double-check this claim, we compute the Lyapunov exponent $\lambda_L$ from semiclassical dynamics~\cite{Chavez2016Classical,LewisSwan2019Scrambling,Skokos2010Book}. A positive $\lambda_L$ indicates exponential divergence of initially adjacent trajectories to signal chaos. The symbols in Fig.~\ref{fig:transition}(d) represent the average $\overline\lambda_L$. They become non-zero when close to $\tilde{g}_c$, but the onset point is noticeably smaller. The lines in the same figure represent $\alpha_{NZ}$, the fraction of sampled trajectories that have $\lambda_L>0.01$. The majority of the trajectories become chaotic (i.e., $\alpha_{NZ}\sim1$) when $\tilde{g}>\tilde g_c$, which is consistent with the complexity transition observed from quantum dynamics. It is apparent that these chaos markers depend on the choices of the energy window size or the cutoff values for $\lambda_L$, and therefore do not yield a single clean-cut $g_c$ value.

\begin{figure}[!bt]
\includegraphics[width=1\columnwidth]{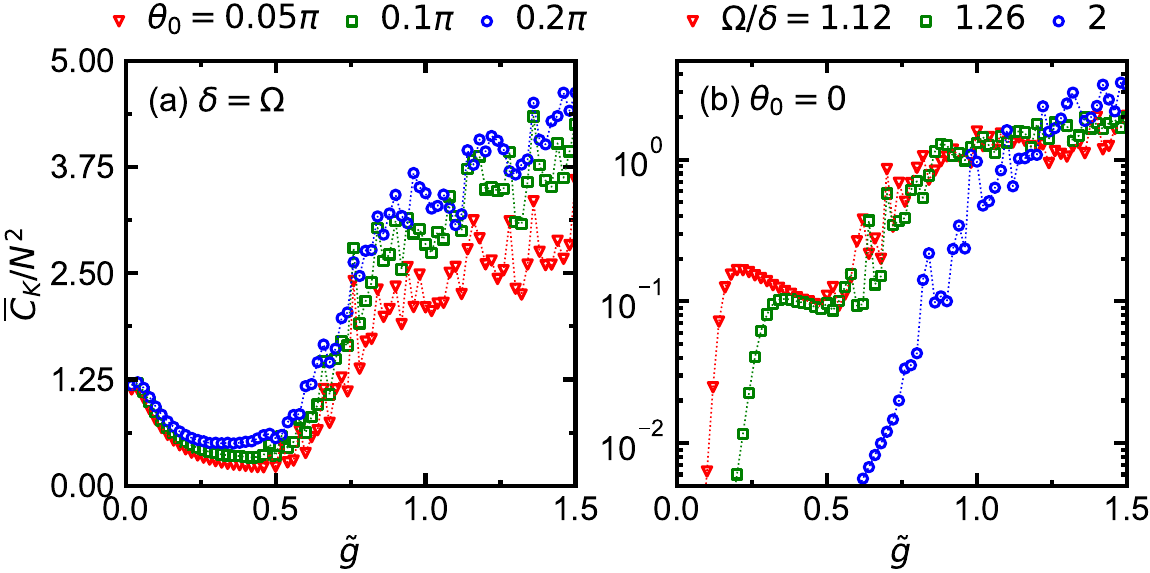}
    \caption{The robustness of the complexity transition. (a) The long-time average $\overline C_K$ for initial states with finite $\theta_0$.(b) $\overline C_K$ (in log scale) off resonance, for a few different $\Omega/\delta$ values. The time average is taken over $gt\in[50,200]$ and $N=20$.}
    \label{fig:otherconditions}
\end{figure}

{\it Persistence of complexity transition}---In the main text, we focused on the quench from the initial state $(\theta_0=0, \alpha_0=0)$ at resonance $\Omega=\delta$. Here, we provide clear evidence that the complexity transition persists to other parameters and is not limited to a few special cases. In Fig.~\ref{fig:otherconditions}(a), we plot the average Krylov complexity $\overline C_K$ for initial angles $\theta_0=0.05\pi$, $0.1\pi$, and $0.2\pi$ at $\delta=\Omega$. Features similar to Fig.~\ref{fig:ckaverage}(a) are observed: In the regular regime, $\overline C_K$ varies smoothly  with $\tilde g$, while in the chaotic regime, $\overline C_K$ increases rapidly to much larger values with more pronounced fluctuations. Fig.~\ref{fig:otherconditions}(b) shows $\overline C_K$ for $\theta_0=0$ but away from resonance. Three horizontal cuts across the complexity phase diagram in Fig.~\ref{fig:ckaverage}(b) are shown for $\Omega/\delta=1.12$, $1.26$, and $2$, respectively. Compared to the resonance case, we find a third regime at small $\tilde g$, where $\overline C_K$ is extremely small and the spins are locked to the initial value, corresponding to the blue color region in Fig.~\ref{fig:ckaverage}(b). Accordingly, the regular regime becomes narrower as $\Omega/\delta$ shifts further away from resonance. In the $\Omega/\delta=2$ case, $\overline C_K$ increases directly from zero to large values indicative of chaotic dynamics, and the regular regime is no longer visible.

\begin{figure}[!bt]
\includegraphics[width=1\columnwidth]{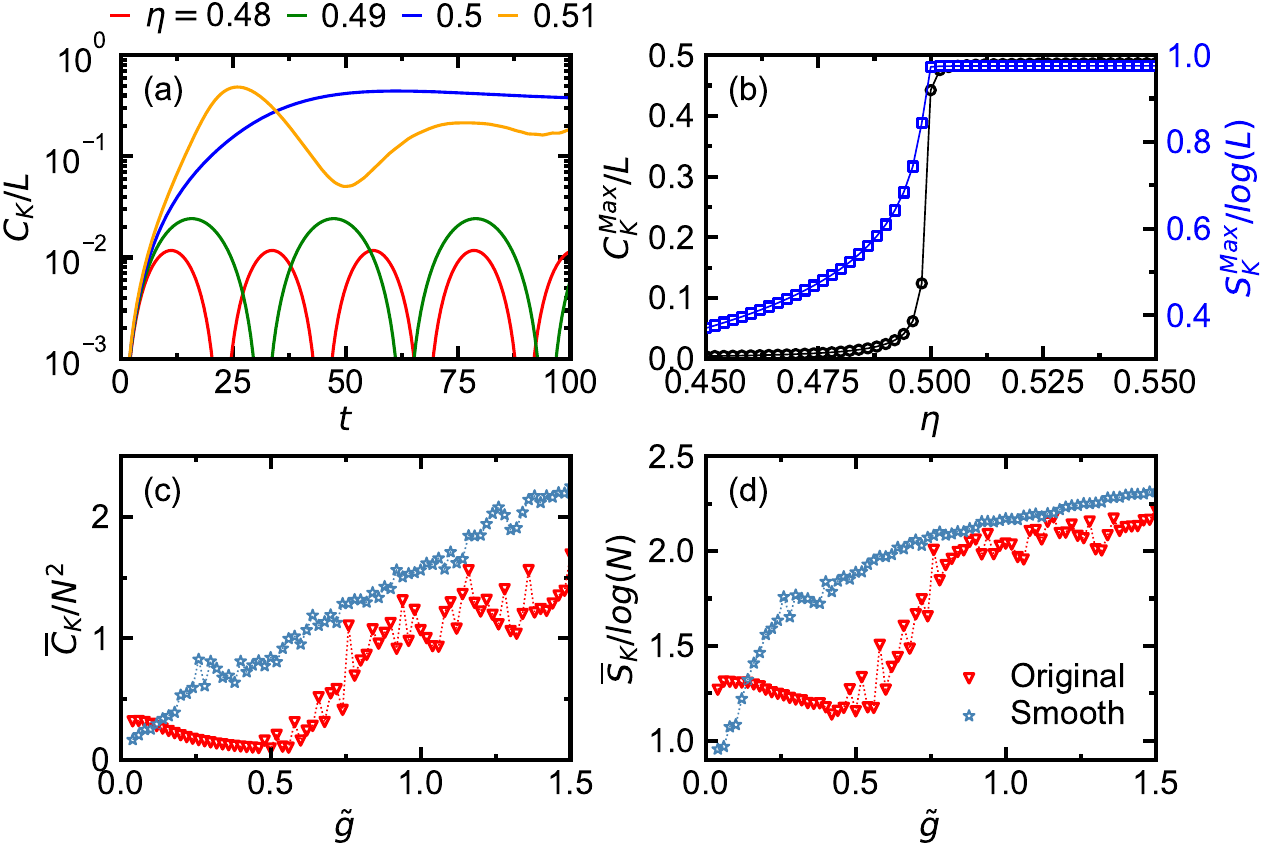}
\caption{(a) $C_K(t)$ for the Rindler-Krylov Hamiltonian Eq.~\eqref{linearab} for a few $\eta$ values near $0.5$. (b) Maximum values of $C_K$ (black) and $S_K$ (blue) as functions of $\eta$ show a clear delocalization transition.$L=1000$ for (a) and (b). (c) $\overline C_K$ of the full model (red triangles) using the original $(a_k,b_k)$ versus the smoothed model (dark blue stars) using $(\overline{a}_k,\overline{b}_k)$.(d) Comparing $\overline S_K$ of the two models. $L=2400$, $N=20$ for (c) and (d).}
    \label{fig:lanczossmooth}
\end{figure}

{\it Wave mechanics of the Rindler-Krylov Hamiltonian}---For convenience, we call the strictly linear model Eq. \eqref{linearab} the Rindler-Krylov (RK) Hamiltonian. Previously, the connection between tight-binding models with $b_k\propto k$ and Rindler spacetime was established through Dirac Hamiltonians near half filling (zero energy)~\cite{rodriguez2017synthetic,morice2021synthetic,morice2022quantum}, and models with power-law $a_k$ and $b_k$ were investigated for their density of states~\cite{moghaddam2021engineering}. Our interest here is different: the nonrelativistic wave dynamics of a single particle without a filled Fermi sea. The initial state localized at the ``horizon" $k=0$ will disperse thanks to its large momentum uncertainty. Moving from site $k$ to $k+1$, the onsite potential increases by $1$, while the local ``band bottom" reduces by $2\eta$. For $\eta<1/2$, the confining potential dominates, the dynamics is oscillatory. The main features of this regime can be captured by perturbation theory. For small $\eta$, the $n$-th eigenstate $\phi_n(k)$ has energy $E_n\sim n$, is centered around $k\sim n$, and has a width $\propto n$. A hallmark of $\phi_n$ is chirping: as it oscillates, the wavelength increases gradually with $k$, due to the confluence of the confining potential (the de Broglie wavelength increases with $k$ according to WKB) and the Rindler hopping $b_k$ that causes ``gravitational redshift." Only a few low-energy eigenstates have considerable overlap with the initial state to be involved in the dynamics. Fig. \ref{fig:lanczossmooth}(a) confirms that $C_K(t)$ oscillates for $\eta<0.5$.

For $\eta>0.5$, the kinetic term $b_k$ wins and the particle prefers to move to the right. This case is not realized in the Dicke model at resonance, where $\eta$ increases monotonically with $\tilde{g}$ to approach $1/2$ from below. A delicate balance is struck at the critical point $\eta_c=1/2$. Assuming open boundary condition at $k=L$, the eigenvalue problem of the RK Hamiltonian leads to recursion relations of the Laguerre polynomials $L_k(2E+1)$, which can be expressed in terms of 
${}_1F_1 (-k;1;2E+1)=\sum_{j=0}^{k}  \binom{k}{j} {(-2E-1)^j}/{j!}$
in the main text. It reduces to the more familiar Bessel function for large $k$ or small $E$, 
$\phi_E(k) \propto (-1)^k J_0(2\sqrt{(2E+1)k})$
which has asymptotic power-law decay and cosine oscillation. The key point is that even the ground state becomes extended throughout the lattice. There are a lot more eigen states with finite overlap with the initial state, in fact the density of states diverges as $1/\sqrt{E}$ at $E=0$. Fig.~\ref{fig:lanczossmooth}(a) shows that $C_K(t)$ rises and saturates at $\sim L/2$ for $\eta=0.5$. Fig.~\ref{fig:lanczossmooth}(b) plots the maximum of $C_K(t)$ and $S_K(t)$ as a function of $\eta$. Their rapid rise to saturation at $\eta=1/2$ marks the delocalization transition, i.e., the spill of the initial Kronecker delta to fill the entire lattice.

{\it Smoothed model misses the complexity transition}---The smoothed model is introduced to flesh out the effects of disorder. We obtain $(\overline{a}_k,\overline{b}_k)$ by averaging the original $a_k$ and $b_k$ over a window $k\pm10$ (for $k\leq 10$, the original values are kept), so they only contain mild site-to-site fluctuations. Note $(\delta a_k, \delta b_k)$ are not random noises. They are generated deterministically by the Lanczos procedure and are referred to as disorder because of their detrimental effects on the wave propagation. Figs. \ref{fig:lanczossmooth}(c)-(d) compare the dynamics of the smoothed model (stars) to that of the original model with disorder (triangles). The smoothed model, while not a bad approximation for $\overline{S}_K$ in the chaotic regime, grossly exaggerates $\overline{C}_K$ in the regular regime and misses the suppression of $\overline{C}_K$ by the disorder and, as a result, the existence of $g_c$. Thus, disorder $(\delta a_k, \delta b_k)$ plays a crucial role in confining the waves in Krylov space. Frame-by-frame comparisons of the wave packet dynamics of the two models are available as movie files in Supplementary Materials~\cite{SM}. It is clear from these movies that while the wave is always ``confined" within some finite region $k\in(0,k_{max})$ for both regimes, the waveform and the meaning of $k_{max}$ change completely as $\tilde{g}_c$ is crossed. In the chaotic regime, the initial wave packet is so elongated that it cannot refocus or exhibit center-of-mass oscillation. It has fallen victim to spaghettification.

\clearpage
\onecolumngrid
\begin{center}
    {\large\bfseries Supplemental Material}\\[3em]
\end{center}

\setcounter{section}{0}
\renewcommand{\thesection}{S\arabic{section}}
\setcounter{subsection}{0}
\renewcommand{\thesubsection}{S\arabic{section}.\arabic{subsection}}
\setcounter{figure}{0}
\renewcommand{\thefigure}{S\arabic{figure}}
\setcounter{table}{0}
\renewcommand{\thetable}{S\arabic{table}}
\setcounter{equation}{0}
\renewcommand{\theequation}{S\arabic{equation}}
\setcounter{page}{1}
\renewcommand{\thepage}{S\arabic{page}}

\twocolumngrid

\section*{Short time dynamics}
\begin{figure}[!bt]
\includegraphics[width=1\columnwidth]{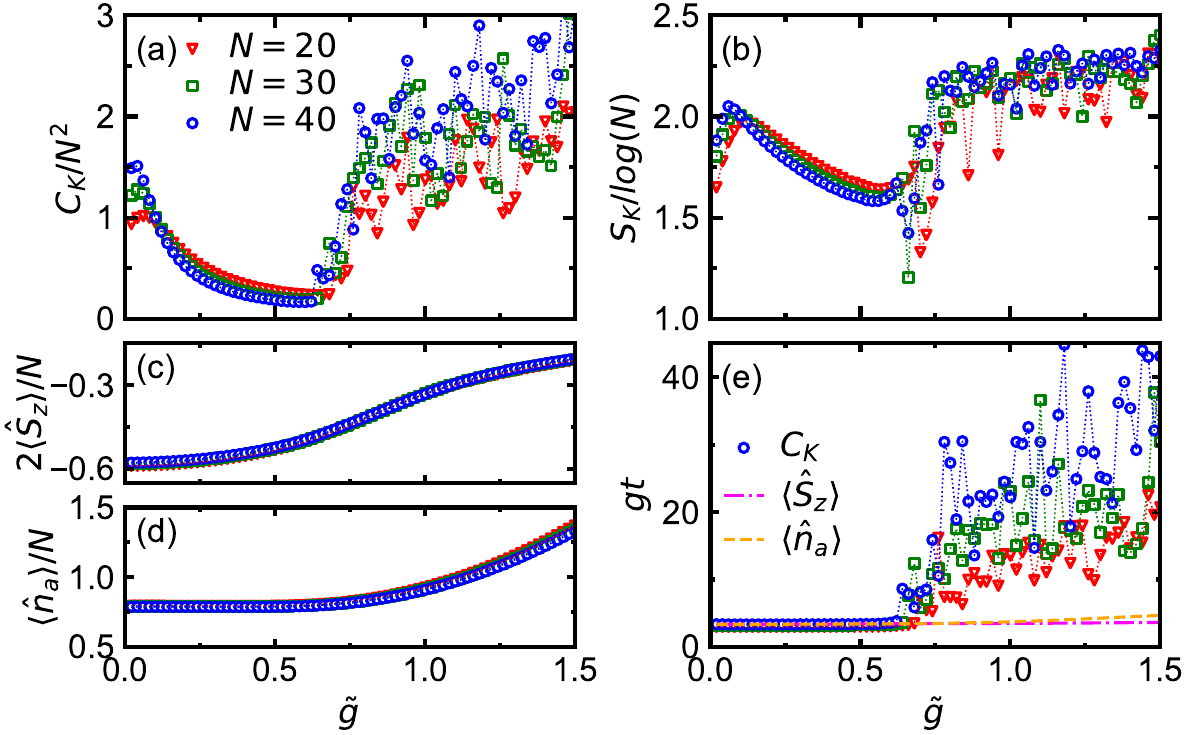}
    \caption{(a) First local maximum of $C_K$ after the quench. Results are shown for $N=20$ (red triangles), 30 (green squares), and 40 (blue circles), with $\delta=\Omega$. (b) Krylov entropy $S_K$ evaluated at the first local maximum of $C_K$. (c) [(d)] First local minimum of $\hat S_z$ (local maximum of $\hat n_a$) after the quench. (e) Symbols: time at which the first local maximum of $C_K$ in (a) occurs. Dash-dotted and dashed lines: corresponding times for the $N=40$ data of $\hat S_z$ and $\hat n_a$ shown in (c) and (d), respectively.}
    \label{fig:shorttime}
\end{figure}

In addition to the long-time average of Krylov complexity, the short time dynamics of $C_K$ already exhibits distinct features in the regular and chaotic regimes. The examples in Fig.~2 (a) and (b) of the main text suggest differences in both the amplitude and location of the first local maximum of $C_K$, which correspond to the oscillation and spaghettification of the wave packet on the Krylov lattice. In Fig.~\ref{fig:shorttime} (a), we plot the first local maximum of $C_K$ after the quench as a function of $\tilde g$. Compared with Fig.~1 (a) of the main text, we observe the same features including the smooth decrease with $\tilde g$ in the regular regime and the rapid increase and strong fluctuations with $\tilde g$ in the chaotic regime. This observation also holds across different system sizes, where a rescaling by $N^2$ suggests a different $N$ dependence in the two regimes.

Fig.~\ref{fig:shorttime} (b) shows the Krylov entropy $S_K$ at the first local maximum of $C_K$. A similar pattern is observed in $S_K$ as a function of $\tilde g$, which follows the behavior of $C_K$ across the transition. In Fig.~\ref{fig:shorttime} (e), we further plot the corresponding time $gt$ of the maximum $C_K$ shown in panel (a). While the maximum $C_K$ changes with $\tilde g$, $gt$ remains roughly constant in the regular regime. This observation suggests similarities in the short-time oscillation of the wave packet on the Krylov lattices, which will be discussed later in this section. On the other hand, we no longer see regular oscillation on the Krylov lattice in the chaotic regime. The first local maximum of $C_K$ occurs at a much later time, and $gt$ shows large fluctuations as a function of $\tilde g$. Note that the transition point obtained from the first-local-maximum analysis occurs at a larger $\tilde g_c$, where the short-time $C_K$ still retains features from the regular dynamics. We thus focused on the long-time average in the main text.

\begin{figure}[!bt]
\includegraphics[width=1\columnwidth]{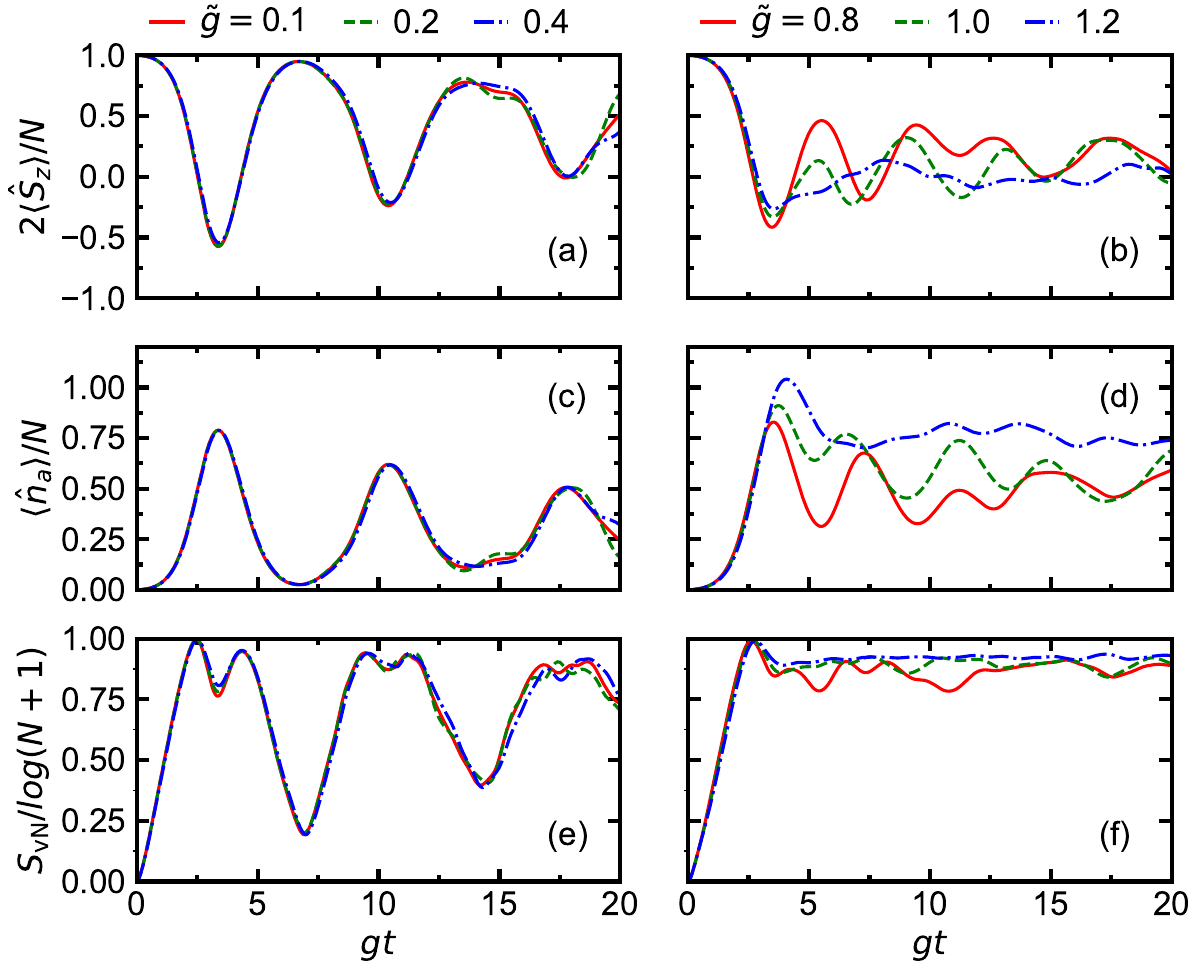}
    \caption{(a), (c), and (e) Short-time dynamics of $S_z$, $n_a$, and $S_{\rm vN}$ for examples in the regular regime. (b), (d), and (f) Same as (a), (c), and (e) but for examples in the chaotic regime. We plot quench dynamics from the $\theta_0=0$ initial state with $\delta=\Omega$ for $N=40$.}
    \label{fig:shortsz}
\end{figure}

For completeness, we show the short time dynamics for other observables that have been used to probe the DPT. In Fig.~\ref{fig:shortsz} (a) and (c), we plot the time evolution of the spin magnetization $S_z$ and boson occupation $n_a$ for examples in the regular regime. We find that results with different $\tilde g$ follow similar nonlinear oscillations to $C_K$, and they almost agree with each other for the first two oscillations we plot. The dips of $S_z$ match the peaks of $n_a$, which reflect the conservation of $S_z+n_a$ in the integrable limit of the Tavis-Cummings model. Fig.~\ref{fig:shortsz} (b) and (d) plot examples of $S_z$ and $n_a$ in the chaotic regime. After the short time decrease of $S_z$ (increase of $n_a$), they undergo erratic oscillations and relax to their equilibrium values. Results for different $\tilde g$ no longer agree with each other. In Fig.~\ref{fig:shorttime} (c) [(d)], we plot the first local minimum of $S_z$ (first local maximum of $n_a$) as a function of $\tilde g$. Unlike $C_K$, these two observables change smoothly with $\tilde g$, with no clear sign of the transition. Fig.~\ref{fig:shorttime} (e) further plots the corresponding times of $S_z$ (dash-dotted line) and $n_a$ (dashed line) in panels (c) and (d). While they agree with the time of the $C_K$ peak in the regular regime, there is no sudden jump in $gt$ across the transition.

Fig.~\ref{fig:shortsz} (e) and (f) plot the spin-boson entanglement entropy $S_{\rm vN}$ for examples in the regular and chaotic regimes, respectively. In both regimes, the spin and boson degrees of freedom become maximally entangled at very short times. For the regular dynamics, the spin and boson degrees of freedom can disentangle, and we observe oscillations in $S_{\rm vN}$. On the other hand, $S_{\rm vN}$ stays at a nearly maximal value for the chaotic dynamics. We see this difference in the averaged $\overline S_{\rm vN}$ in Fig.~3(b) of the End Matter. 

\begin{figure}[!bt]
\includegraphics[width=1\columnwidth]{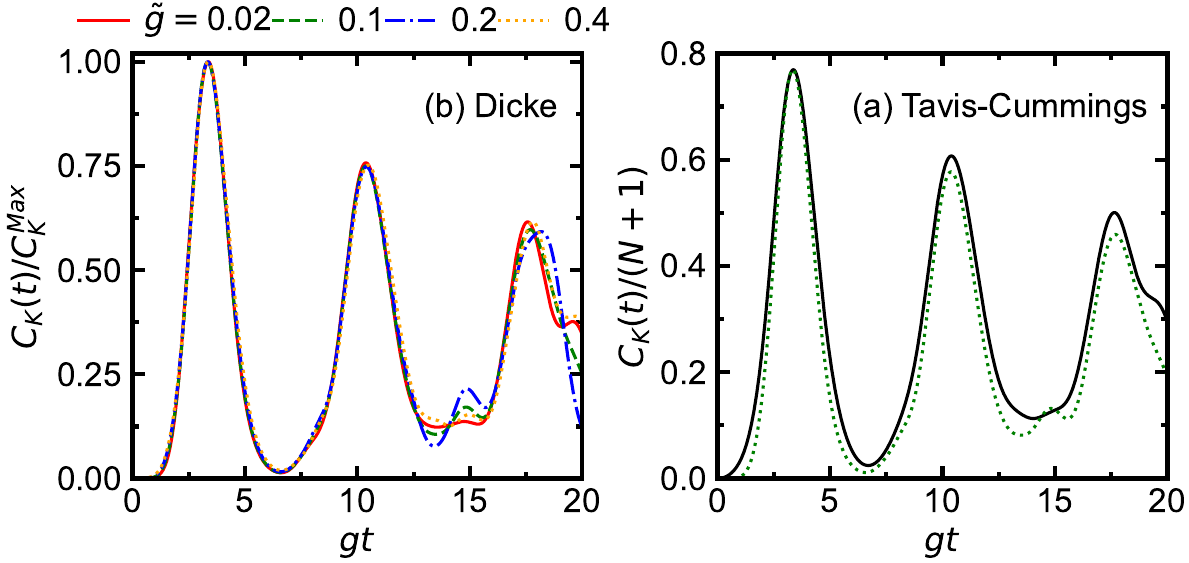}
    \caption{(a) Dynamics of $C_K$ rescaled by its maximum value $C_K^{\rm Max}$ (within $gt\in[0,200]$) for examples in the regular regime of the Dicke model. We show results for a quench from the initial state $(\theta_0,\phi_0,\alpha_0)=(0,0,0)$. (b) $C_K$ in the Tavis-Cummings model normalized by the Krylov space dimension $N+1$ (black solid line). The dotted line is for the corresponding quench in the Dicke model at $\tilde g=0.1$, with a scaling that matches the first peak of the data. $N=40$ in this plot.}
    \label{fig:CkTC}
\end{figure}

Since the first $C_K$ peak in the regular regime occurs at the same location, we further plot the short-time dynamics of the rescaled Krylov complexity $C_K/C_K^{\rm Max}$ in Fig.~\ref{fig:CkTC}(a), where $C_K^{\rm Max}$ is the global maximum for $gt\in[0,200]$. We find that all $C_K$ curves have a similar short-time profile, differing primarily in their amplitudes. Moreover, the global maximum $C_K^{\rm Max}$ occurs at the first peak. This result suggests that there are similar structures in the Krylov lattices for the regular regime dynamics. The observation in $C_K$ is also consistent with the short time dynamics of other observables (see Fig.~\ref{fig:shortsz}), where the shape is the same and there is no difference in amplitude.

\section*{Tavis-Cummings limit}
To further understand the dynamics of $C_K$ in the regular regime, we study the integrable limit of the Tavis-Cummings model. For a state in the Fock basis $|S_z,n_b\rangle$, applying the Tavis-Cummings Hamiltonian $\hat H_{\rm TC}=-({g}/{\sqrt{N}})(\hat a\hat S^++\hat a^{\dagger}\hat S^-)$ gives
\begin{align}
    &\hat H_{\rm TC}|S_z,n_b\rangle\nonumber=-\frac{g}{\sqrt{N}}\\&\bigg[\sqrt{n_b+1}\sqrt{\frac{N}{2}(\frac{N}{2}+1)-S_z(S_z-1)}|S_z-1,n_b+1\rangle \nonumber\\&+\sqrt{n_b}\sqrt{\frac{N}{2}(\frac{N}{2}+1)-S_z(S_z+1)}|S_z+1,n_b-1\rangle\bigg],\nonumber
\end{align}
which has the same tight-binding form as that on the Krylov lattice [main text Eq.~(2)]. Thus, for our quench initial state $|\psi_0\rangle=|N/2,0\rangle$, the Krylov basis is the same as the Fock basis, and the dynamics of $C_K$ shares the same features with $\langle\hat S_z\rangle$ up to some constants. Note that the connection between $C_K$ and $\langle\hat S_z\rangle$ also exists for quench dynamics in the LMG model~\cite{Bento2024Krylov}.

Fig.~\ref{fig:CkTC}(b) plots the short-time dynamics of $C_K/(N+1)$ from the initial state $|\psi_0\rangle=|N/2,0\rangle$ in the Tavis-Cummings model (solid line). Comparing with the Dicke model in the regular regime (dotted line; example for $\tilde g=0.1$ divided by an extra factor of 50.8), we find oscillations and the same peak locations in $C_K$ but with much smaller values. The matching of oscillation period and peak locations is expected, since the dynamics of $\langle\hat S_z\rangle$ in the Tavis-Cummings model is the same as that of $C_K$, and the short-time dynamics of $\langle\hat S_z\rangle$ is very close to that of the Dicke model at small $\tilde g$. The shape of the $C_K$ peak in the Dicke model is narrower, which suggests a faster spread of the wave packet in the Krylov lattice when the number of boson excitations is larger.

\begin{figure}[!bt]
\includegraphics[width=1\columnwidth]{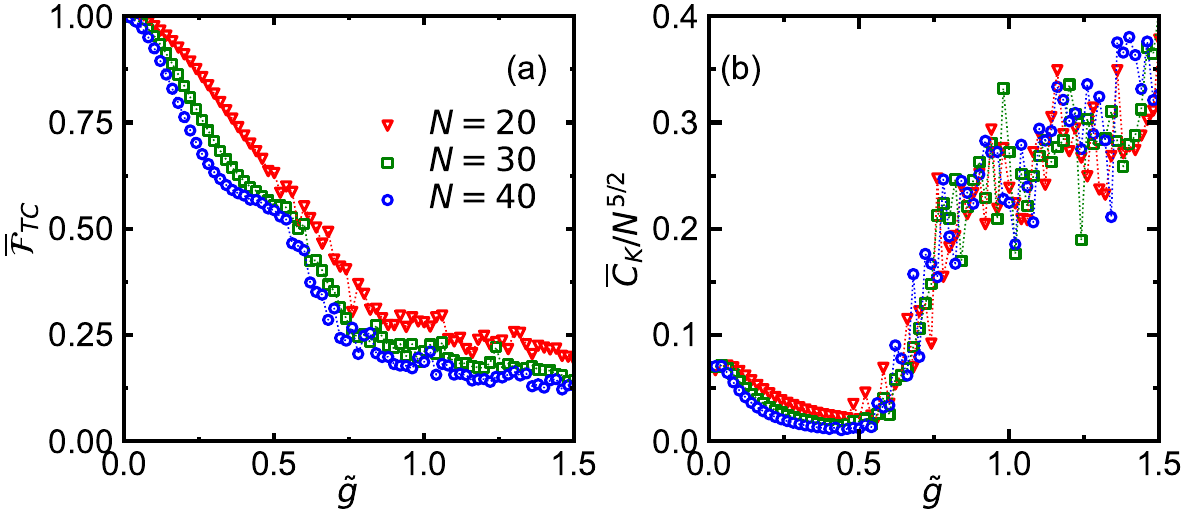}
    \caption{(a) Time-averaged fidelity $\overline{\cal F}_{\rm TC}$ for the projection of$|\psi(t)\rangle$ onto the sub-Hilbert space of the Tavis-Cummings model. (b) Time-averaged $\overline C_K$ rescaled by $N^{5/2}$. We show quench results from the $\theta_0=0$ initial state and set $\delta=\Omega$. The time average is taken over $gt\in[50,200]$.}
    \label{fig:TCoverlap}
\end{figure}

We finish this section by studying the projection of the Dicke model wavefunction $|\psi(t)\rangle$ onto the subspace of the Tavis-Cummings Hamiltonian dynamics $\{|K^{\rm TC}_n\rangle, n=0,1,\cdots,N\}$, ${\cal F}_{\rm TC}=\sum_n|\langle K^{\rm TC}_n|\psi(t)\rangle|^2$. In Fig.~\ref{fig:TCoverlap} (a), we plot the time-averaged $\overline{\cal F}_{\rm TC}$. While $\overline{\cal F}_{\rm TC}$ roughly follows a smooth decrease with $\tilde g$ within both regimes, we observe an inflection point at the transition. The inflection point suggests that there are qualitative changes in the post-quench wavefunction at the transition, and it becomes more significant as $N$ increases. Note that while $\overline C_K$ shows a peak at very small $\tilde g$ [see, e.g., main text Fig.~1(a)], we do not see this in ${\cal F}_{\rm TC}$.

\section*{Scaling of $\overline C_K$}
In Fig.~1 (a) and (c) of the main text, we show the scaling of $\overline C_K$ by $N^2$ and $N^{3/2}$. While neither of them fully captures the $N$ dependence across the transition, we try another scaling $\overline C_K/N^{5/2}$ in this section. As shown in Fig.~\ref{fig:TCoverlap}(b), the extra $1/\sqrt{N}$ factor makes the data agree better in the chaotic regime. The different $N^\alpha$ scaling in the two regimes follows our expectation: the regular regime retains some features of the extra conservation relations in the integrable limit, at which the size of the Krylov space is limited to $N+1$ for our quench. Without this constraint, the chaotic dynamics spans the Hilbert space of $N+1$ spin states times the number of boson states ($\sim N$), and we expect to see a larger $\alpha$. Interestingly, we notice that the peak of $\overline C_K$ at very small $\tilde g$ also roughly scales as $N^{5/2}$. We leave the following questions for future work: (1) the values of scaling coefficients $\alpha$ in different regimes and their universality; (2) the unexpected large $\overline C_K$ values for small $\tilde g$ and their scaling.

\begin{figure}[!bt]
\includegraphics[width=1\columnwidth]{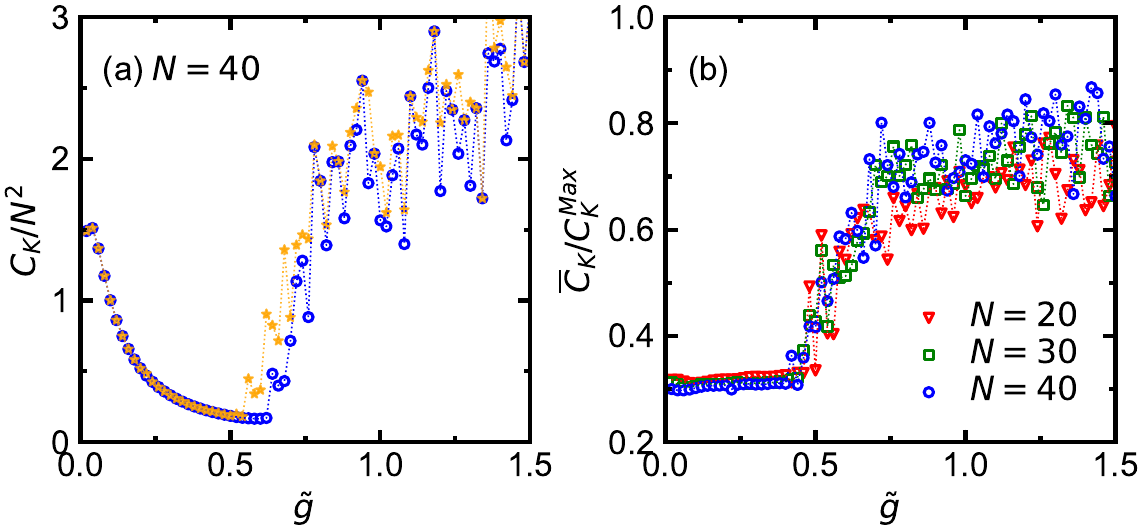}
    \caption{(a) First (blue circles) and global (orange stars) maxima of $C_K/N^2$ within $gt\in[0,200]$ for $N=40$ as a function of $\tilde g$ at $\delta=\Omega$. (b) Ratio of the long-time average to the maximum, $\overline C_K/C_K^{\rm Max}$. $\overline C_K$ is averaged over $gt\in[50,200]$, and $C_K^{\rm Max}$ is computed within $gt\in[0,200]$.}
    \label{fig:ckmax}
\end{figure}

\section*{Maximum of $C_K$}
In addition to the long-time plateau, the peak value of $C_K$ has been proposed to characterize the integrable-chaotic transition in the literature \cite{Baggioli2025Krylov}. In Fig.~\ref{fig:ckmax} (a), we plot the global maximum of $C_K$ as a function of $\tilde g$ (orange stars) for a system of $N=40$ at $\delta=\Omega$. We observe behavior in $C^{\rm Max}_K$ similar to the long-time average $\overline C_K$, and there is a sudden jump in $C^{\rm Max}_K$ across the transition. As a reference, we plot the corresponding first local maximum of $C_K$ [blue circles, see also Fig.~\ref{fig:shorttime}(a)]. Inside the regular regime, the first local maximum agrees with the global one. In the chaotic regime, both quantities have large values with strong fluctuations as a function of $\tilde g$. The most significant difference occurs on the chaotic side near the transition, where the first local maximum remains a smooth function of $\tilde g$ while the global maximum starts to increase. The transition point in the global maximum of $C_K$ is consistent with the observation from the time-averaged quantities.
Fig.~\ref{fig:ckmax} (b) plots the ratio between the long-time average and the maximum of $C_K$. In the regular regime, $\overline C_K/C_K^{\rm Max}$ roughly stays at a small constant value as $\tilde g$ varies, as a result of the robust oscillation dynamics of the wave packet on the Krylov lattice. The ratio rapidly rises to a larger value ($\sim0.8$) with the spaghettification of wave packet in the chaotic regime.

\section*{LMG limit}
In the limit $\Omega\ll\delta$, the Dicke Hamiltonian [Eq.~(1) of main text] reduces to the integrable spin LMG Hamiltonian $$\hat H_{\rm LMG}=-\frac{\chi}{N}\hat S^2_x+\Omega \hat S_{z}\,,$$where $\chi\equiv4g^2/\delta$. Similar to the Tavis-Cummings limit discussed in the main text, the size of the Krylov space is limited to $N+1$. As an interesting example, we will discuss the quench from the initial state $(\theta_0,\phi_0,\alpha_0)=(\pi/2,0,0)$. In the LMG limit $\Omega/\delta\to0$, there is a transition from an untrapped to a trapped integrable phase as $\tilde g$ increases, with a critical value $\tilde g_c=\sqrt{2}$. As $\Omega/\delta$ increases, there exist regimes with chaotic dynamics \cite{LewisSwan2021Characterizing}.

\begin{figure}[!bt]
\includegraphics[width=1\columnwidth]{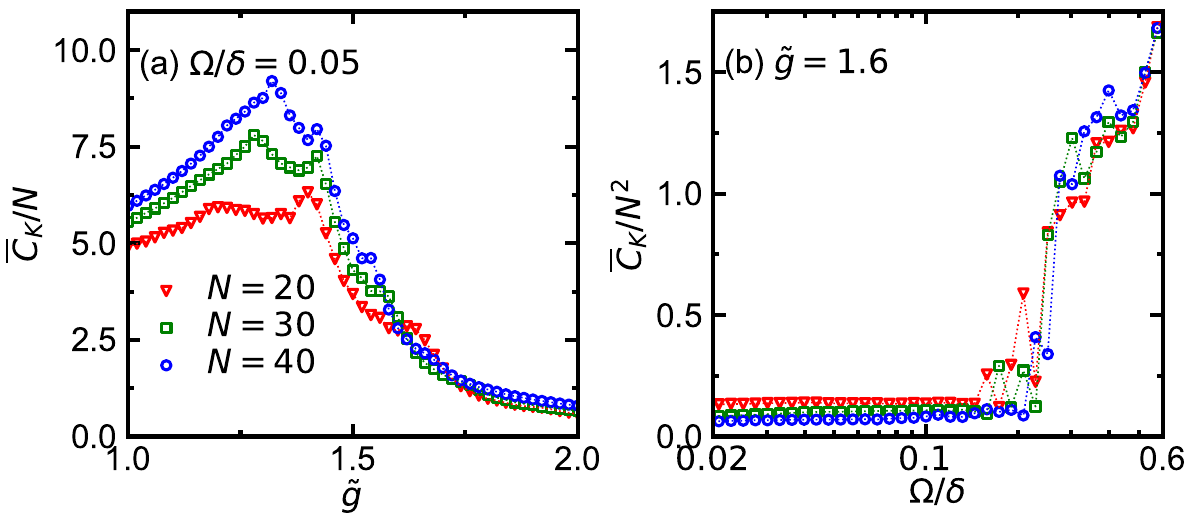}
    \caption{(a) The long-time average $\overline C_K/N$ near the LMG limit $\Omega/\delta\ll1$. We fix $\Omega/\delta=0.05$ and vary $\tilde g$. The initial state is chosen as $\theta_0=\pi/2$, and the time average is taken over $gt\in[50,200]$. (b) $\overline C_K/N^2$ as a function of $\Omega/\delta$ for fixed $\tilde g=1.6$.}
    \label{fig:LMG}
\end{figure}

Fig.~\ref{fig:LMG} (a) plots the normalized $\overline C_K/N$ as a function of $\tilde g$ at $\Omega/\delta=0.05$. The transition from the untrapped to the trapped phase can be seen in $\overline C_K$, with a cusp at the transition point $\tilde g_c$. In the untrapped phase, where the dynamics spans the entire Bloch sphere, we observe a relatively larger $\overline C_K$; while in the phase where the spin is trapped near its initial state, $\overline C_K$ is relatively smaller. Results for different $N$ suggest that the scaling coefficient $\alpha>1$ for $N^\alpha$. Fig.~\ref{fig:LMG} (b) plots $\overline C_K/N^2$ as a function of $\Omega/\delta$ at $\tilde g=1.6$. We see the jump in $\overline C_K$ when the dynamics changes from integrable to chaotic. Across the transition, we notice a different scaling $\overline C_K\sim N^\alpha$, where $\alpha<2$ for the integrable dynamics. Thus, our observations here are consistent with Refs.~\cite{LewisSwan2021Characterizing, Bento2024Krylov}. 
These examples provide further evidence that Krylov complexity can be used as a general, effective tool to identify and characterize qualitatively different regimes of quench dynamics.

\section*{Linear model of Lanczos coefficients}

\begin{figure}[!bt]
\includegraphics[width=1\columnwidth]{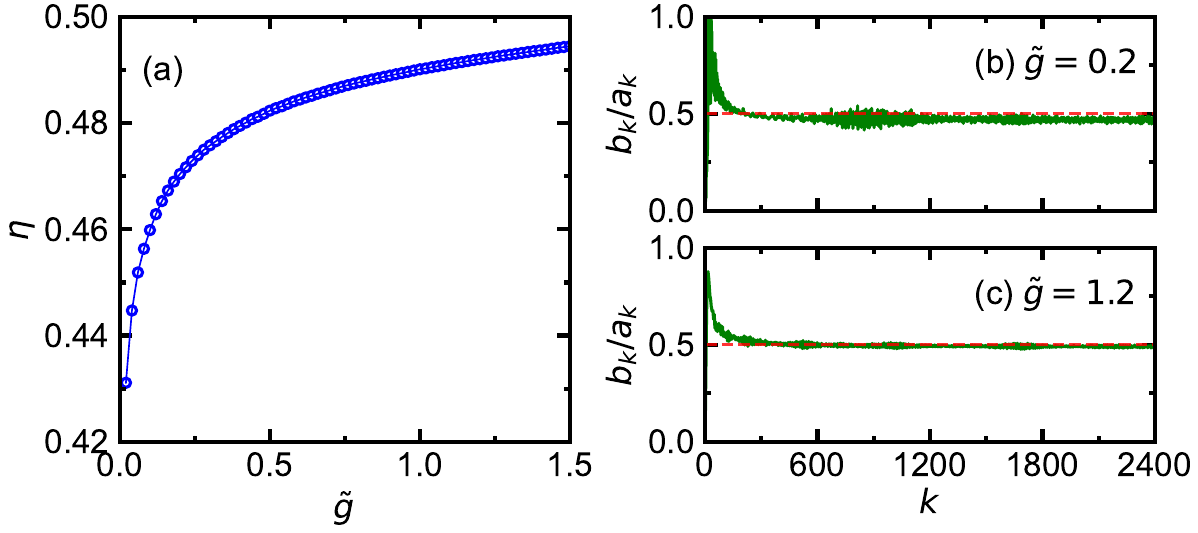}
    \caption{(a) We fit the Lanczos coefficients $a_k$ and $b_k$ calculated for the Dicke model quench to the Rindler-Krylov Hamiltonian in Eq.~(3) of the main text. The slope ratio $\eta$ is plotted as a function of $\tilde g$ for quenches from the $\theta_0=0$ initial state at $N=20$ and $\delta=\Omega$. The fitting range is $k\leq 2400$. (b) and (c) Examples of $b_k/a_k$ for $\tilde g=0.2$ in the regular regime and $\tilde g=1.2$ in the chaotic regime. Red dashed lines mark $b_k/a_k=0.5$ as a guide to the eye.}
    \label{fig:linearslope}
\end{figure}

We analyze the Lanczos coefficients calculated for the Dicke-model quench by fitting them to the toy model of the Rindler-Krylov Hamiltonian introduced in Eq.~(3) of the main text. In Fig.~\ref{fig:linearslope} (a), we plot the ratio of the slopes $\eta=b_k/a_k$ extracted from the fit as a function of $\tilde g$. Results show that $\eta$ increases monotonically with $\tilde g$, approaching the critical point $\eta=0.5$ from the localized side. There is no hint of the $\overline C_K$ transition from the regular to the chaotic regime in this toy model.
Fig.~\ref{fig:linearslope} (b) and (c) further show examples of $b_k/a_k$ as a function of $k$ in the regular and chaotic regimes, respectively. For small $k$, we see that the ratios are greater than 0.5, which is in agreement with our picture that the wave packet gets `launched' with a finite momentum. For large enough $k$, $b_k/a_k$ roughly stays at a constant value below $0.5$, which confines the wave packet in the Krylov lattice. The main difference between these two examples is the `disorder' $\delta a_k$ and $\delta b_k$. As shown in the End Matter, they are crucial to the transition in $\overline C_K$.

\section*{Wave packet dynamics}
We provide movie files that illustrate the wave-packet dynamics $p_k(t)$ in Krylov space for the Dicke model on resonance, with $\tilde{g}=0.2$ to $1.2$ in steps of 0.2. The upper panels in the movies are obtained from the original $(a_k,b_k)$; the lower panels are obtained using $(\bar{a}_k,\bar{b}_k)$.

\end{document}